\documentclass[allclo]{FBSart}
\usepackage{amsfonts}
\usepackage{amssymb}
\usepackage[english]{babel}
\usepackage{graphicx}  
\usepackage{bm}  
\usepackage{amsmath}
\usepackage{epsf}
\newcommand{\beq}{\begin{equation}}
\newcommand{\eeq}{\end{equation}}
\newcommand{\bea}{\begin{eqnarray}}
\newcommand{\eea}{\end{eqnarray}}       
\newcommand{\dd}{\mathrm{d}}

\newcommand{\kvec}{{\bf k}}

\newcommand{\qvec}{{\bf q}}
\newcommand{\pvec}{{\bf p}}

\def\vec#1{{\bf #1}}

\begin{document}
\title{Low-Energy Universality\\
in Atomic and Nuclear Physics}
\runningtitle{Low-Energy Universality in Atomic and Nuclear Physics}
\author{Lucas Platter}
\runningauthor{L.\ Platter}
\institute{Department of Physics,
Ohio State University, Columbus, OH 43210, USA\\}
\date{\today}
\maketitle
\begin{abstract}
  An effective field theory developed for systems interacting through
  short-range interactions can be applied to systems of cold atoms with a
  large scattering length and to nucleons at low energies. It is therefore the
  ideal tool to analyze the universal properties associated with the Efimov
  effect in three- and four-body systems.  In this {\it progress report}, we
  will discuss recent results obtained within this framework and report on
  progress regarding the inclusion of higher order corrections associated with
  the finite range of the underlying interaction.
\end{abstract}
\newpage
\tableofcontents
\newpage
%
%
\section{Introduction}
\label{sec:intro}
Universal properties are one reason why few-body systems with a large
scattering length $a$ have received a lot of attention lately. For example, a
two-body system with a large positive scattering length $a$ has a bound state
with binding energy
\begin{equation}
  E_D=\frac{1}{m a^2}~,
\end{equation}
where $m$ denotes the mass of the particles (note that we set $\hbar=1$ for
convenience). In the three-body sector a large two-body scattering length
leads to further universal features known widely as Efimov
physics~\cite{Efimov70}. These are fascinating and exciting since they relate
physical effects that occur at very different length scales. In experiments
with ultracold atoms Feshbach resonances allow the interatom scattering length
to be tuned from $-\infty$ to $\infty$, while also the internucleon interaction
exhibits a scattering length large compared to the range of the interaction.

An understanding of such systems improves therefore also our understanding of
low-energy nuclear physics. Moreover, a simple approach only in terms of the
large scattering and the other remaining low-energy parameters of the
effective range expansion grants also the possibility to address important
problems in nuclear few-body systems in a very effective manner. One prominent
example is the model-independent description of electroweak reactions
relevant to nuclear astrophysics. An approach whose parameters are only
related to the effective range expansion might seem inappropriate for such
nuclear reactions but it is important to realize that many important processes
occur at energies well below the energy scale set by the range of the
interaction (i.e. the pion mass $m_\pi$).

Since such an approach will fail for observables for which the pion-exchange
character of the internucleon interaction is important, it can therefore also
answer the question for which measurable quantities the chiral character of
the nucleon-nucleon interaction is relevant. This addresses the question what
low-energy information in particular is required to describe nuclear systems
at different energy scales.

The framework that we will use to address processes in such systems is that
of effective field theory (EFT). It allows to obtain model-independent results
in a small parameter expansion. The advantages of a calculation of observables
in the EFT framework are numerous. Firstly, it allows for a clear separation
of the unknown ultraviolet and known low-energy physics which eliminates any
possible model-dependence. Secondly, the accuracy of calculated observables
can be systematically improved by calculation of another order in the
low-energy expansion. And thirdly, the quantum field-theoretic setup brings
along the advantages of mathematically tools successfully used in high-energy
physics, such as Feynman diagrams, renormalization and regularization.

Here we will concentrate on applications of one particular EFT (that we will
call {\it short-range EFT}) that is tailored to the problem of short-range
interactions. The expansion parameters in this EFT are $R/a$ and $k R$, where
$R$ denotes the range of the underlying interaction and $k$ the momentum scale
of the process under consideration. At leading order, this EFT framework allows
us to analyze the properties of systems interacting through zero-range
interactions. Finite range effects can then be accounted for by calculating
subleading orders in the EFT expansion.

The short-range EFT provides therefore a good framework to analyze the
few-body dynamics in resonantly interacting ultracold gases. Binding energies
and scattering observables can be calculated and the experimental signals of
Efimov physics can therefore be identified.

Predictions for low-energy scattering processes and binding energies can also
be made in the nuclear case. However, the field-theoretic formulation of this
framework facilitates in addition a calculation of electroweak observables
relevant to nuclear astrophysics as discussed above.

The purpose of this review is to display the similarities between low-energy
atomic and nuclear physics and to demonstrate that the same framework can be
applied to both sectors to obtain relevant and interesting results.  In
contradistinction to recent reviews on similar topics
\cite{Braaten:2004rn,Braaten:2006vd,Bedaque:2002mn,Epelbaum:2008ga}, we will
present the applications and results for atomic and nuclear systems
alongside. By doing this we want to emphasize the similarities between
low-energy nuclear physics and the few-body dynamics in resonantly interacting
ultracold gases of alkali atoms.

We will be mainly interested in few-body systems and will therefore mainly
focus on systems that can be described either as two, three or four-body
systems. It should also be noted that this is not a complete review of work
done for systems with short-range interactions but instead it should be
considered as a {\it progress report} that gives a brief introduction to the
short-range EFT, that presents recent calculations performed in the few-body
sector and discusses then an incomplete list of problems that should be
addressed in the near future. 

In the following chapter, we will thus give a brief introduction to the
concept of EFTs and will discuss in detail the short-range EFT whose
application to different problems will be covered in the remaining sections of
this review. In section \ref{sec:amo}, we will give a review of recent
applications of the short-range EFT to systems of ultracold atoms.  In section
\ref{sec:nuc} we will review results obtained with short-range EFT for nuclear
systems such as the three- and four-nucleon systems but also heavier systems
such as halo nuclei.  We will conclude with a summary.

\section{Effective Field Theories}
\label{sec:eft}
Effective field theories are an excellent tool to describe the interactions in
systems with a separation of scales. The usual example of such a scale
separation is the ratio of a small momentum over a heavy mass scale. The weak
interaction for example, can be described without explicit use of heavy vector
bosons if the energies of the processes are well below the mass scale of the
exchange particles. In this case the weak interactions can be described very
well with the familiar Fermi interaction.

Here, we will focus on the short-range EFT, which is tailored to describe
non-relativistic particles with a large scattering length $a$ at low
energies. We will discuss the Lagrangian which is at the heart of the EFT and
describe how it can be used along with an ordering scheme called power counting
and Feynman rules to obtain a powerful framework for the computation of
low-energy observables of few-body systems with a large two-body scattering
length.

Very good introductions to the general concepts of EFTs have been written by
Kaplan \cite{Kaplan:2005es}, Lepage \cite{Lepage:1997cs} and Polchinski
\cite{Polchinski:1992ed}. An excellent introduction to EFTs applied to nuclear
systems has been given by Phillips \cite{Phillips:2002da}.
\subsection{The  Effective Range Expansion}
We will consider identical particles (bosons for simplicity) interacting via a
potential that has a finite range $R$. In this case the amplitude for the
scattering of two particles can be written as
\begin{equation}
  \label{eq:tmatrix}
  \mathcal{A}(k,\cos \theta)=\frac{8 \pi}{m}\sum_l \frac{2l +1}{k \cot \delta_l -i k}P_l (\cos \theta)~,
\end{equation}
where $k$ denotes the relative momentum between the two particles, $m$ is
the mass of each of these particles and $\delta_l$ denotes the scattering
phase shift in partial wave $l$.

At low momenta, the expression in the denominator can be expanded in even
powers of the momentum $k$
\begin{equation}
k^{2l+1} \cot\delta_l=-\frac{1}{a_l}+\frac{r_l}{2}k^2+\mathcal{O}(k^4)~.
\end{equation}
For the $S$- and $P$-wave phase shifts, the first terms in the expansions
are
\begin{equation}
k \,\cot \delta_0 =-\frac{1}{a}+\frac{1}{2}r_s k^2 +...~,\:\:\:k \,\cot \delta_1 =-\frac{3}{k^2 a_p^3}+...~,
\end{equation}
where $a$, $ a_p$ are the $S$- and $P$-wave scattering lengths, respectively,
$r_s$ is the $S$-wave effective range and the ellipses denote higher order
terms in the expansion.

For large positive scattering length, one finds a bound state as discussed
above and we will frequently use the effective range expansion (ERE) around the
bound state pole:
\begin{equation}
  \label{eq:ere-pole}
  k \,\cot \delta_0 =-\gamma+\frac{\rho}{2}(\gamma^2+k^2)+...~,
\end{equation}
where $E_D=\gamma^2/m$ and $\rho=r_s$ up to the orders considered in this work.

The low-energy physics encoded in the effective range parameters depends on
the relative size of these parameters. In the following we will consider
two special cases which we will refer to as the {\it natural case} and the
{\it unnatural case}. In the natural case, all effective range parameters are
of natural size, i.e.  of the order of the range of the interaction $a \sim
r_s \sim R$. In this case the scattering amplitude in Eq.~(\ref{eq:tmatrix})
can be expanded in powers of $k$:
\begin{equation}
\label{eq:smalla}
\mathcal{A}(k)=-\frac{8 \pi a_s}{m}\left[1-\hbox{i}a_s k+(a_s
  r_s/2-a_s^2)k^2+\ldots\right]~.
\end{equation}
For simplicity we have considered here only the $S$-wave projected part of the
scattering amplitude. This expansion will converge for momenta $k \ll 1/R$~.

Now consider the case in which the scattering length $a$ is much larger than
all other effective range parameters $|a|\gg |r_s|\sim R$. This separation of
scales indicates the presence of non-perturbative physics through a
(virtual) bound state with energy $\sim 1/(ma^2)$. The
expansion above is only valid for $a k \ll 1$ which reduces the radius of
convergence dramatically and renders it useless in the limit of infinite
scattering length. We will instead expand in powers of $k R$ and keep $a k$ to
all orders:
\begin{equation}
  \label{eq:largea}
  \mathcal{A}(k)=-\frac{8\pi}{m}\frac{1}{1/a+i k}\left[1+\frac{r_s/2 }{1/a+i k}k^2
+\frac{(r_s/2)^2}{(1/a+i k)^2}k^4+\ldots\right]~.
\end{equation}
The resulting expansion still converges for $k R\ll 1$ and reflects the fact that
systems with a large two-body scattering length require a non-perturbative
resummation at {\it leading order}. 

\subsection{The Short-Range EFT} 
Our goal is to describe the dynamics of particles interacting through a
short-range potential with an EFT. The Lagrange density for an EFT is
generated by writing down all possible operators built from the available
degrees of freedom in accordance with the required symmetries.  At
sufficiently low momenta, non-relativistic particles can be described by an
EFT built up from contact interactions alone. This EFT can be applied if the
relative momenta $k$ of the particles is much smaller than the inverse of the
range of the underlying interaction $R$:
\begin{equation}
k \ll 1/R~.
\end{equation}
The convergence radius of our EFT will therefore agree with that of the ERE
and we can therefore anticipate that we will be able fix any free parameters
in the two-body sector from the ERE.

The only degrees of freedom we require for our EFT are therefore
the atoms themselves. Having identified the relevant degrees of freedom,
we generate the effective Lagrangian by writing down all possible operators
satisfying the constraints of Galilean, parity and time-reversal invariance
and locality:
\begin{eqnarray}
\nonumber
  \label{eq:Lagrange1}
\mathcal{L}&=&{\psi}^{\dagger}\left[i
  \partial_t+\frac{{\nabla}^2}{2m}\right]\psi
-\frac{C_0}{4}(\psi^\dagger\psi)^2
-\frac{D_0}{36}(\psi
^{\dagger}\psi)^3-\frac{E_0}{576}(\psi
^{\dagger}\psi)^4...,
\end{eqnarray}
The ellipses represent operators of higher
dimension which means terms with more derivatives and/or more fields. We have
neglected relativistic effects which are suppressed by factors of
$(p/M)^2$. $D_0$ and $E_0$ denote the leading three- and four-body
interactions.

Every EFT contains therefore an infinite number of two- and many-body
operators. This framework might therefore seem useless since an infinite
number of Feynman diagrams arises for every observable and infinite number of
coupling constants needs to be determined from experiment. Here, the
afore-mentioned separation of scales allows us to resolve this problem.  The
separation of scales usually leads to a ratio of a light scale over a heavy
scale (e.g. a small momentum divided by a large mass) which is exploited as a
small expansion parameter $\alpha_{\rm EFT}$. Appropriate renormalization of
the vertex constants (also known as low-energy constants) will then lead to an
ordering of all possible Feynman diagrams (power counting) such that only a
well-defined number of low-energy constants are required at each order of the
expansion of a matrix element in powers of $\alpha_{\rm EFT}$. This
power counting arises in the short-range EFT by demanding that the scattering
amplitude for two-body scattering reproduces the appropriate momentum
expansion (Eqs.~(\ref{eq:smalla}) and (\ref{eq:largea})) of the scattering amplitude.

\begin{figure}[t]
\centerline{\includegraphics*[width=4.in,angle=0]{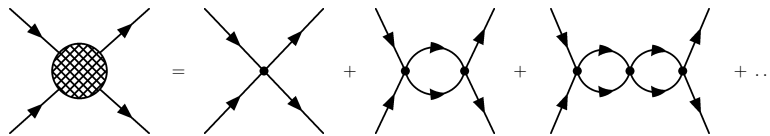}}
  \caption{\label{fig:twobody}
The sum of diagrams including the $C_0$ vertex contributing to
two-body scattering.
}
\end{figure}

\paragraph{\it Feynman rules:\\}
As in standard quantum field theory, amplitudes for processes are
calculated by using Feynman rules derived from the Lagrange density.
The Feynman rules for the short-range EFT are particularly simple
\begin{itemize}
\item Assign non-relativistic four-momenta $(p_0,\pvec)$ to all lines and
  enforce momentum conservation at each vertex.
\item For each vertex include a factor $-i$ times the
low-energy constant of the corresponding operator.
\item For each internal line include the propagator with four-momentum
$(p_0,\pvec)$
\begin{equation}
  \label{eq:propagator}
  i S({\bf q},q_0)=\frac{i}{q_0-\frac{q^2}{2 m}+i \epsilon}~,
\end{equation}
\item Integrate over all undetermined loop momentum using the measure
\beq
\nonumber 
\int\frac{\hbox{d}q_0}{2\pi}\frac{\hbox{d}^3q}{(2\pi)^3}~.
\eeq
The energy integral can be evaluated using contour integration.
\item Multiply by a symmetry factor $1/n!$ if the diagram is invariant under
  the permutation of $n$ internal lines.
\end{itemize}
\paragraph{\it Power counting\\}
It is clear from the discussion on the expansion of the two-body amplitude for
large scattering length that the power counting requires a resummation of an
infinite set of diagrams to obtain the pole structure of
Eq.~(\ref{eq:largea}).  Here, we will verify by explicit calculation that
the leading order two-body amplitude can be reproduced by summing up all
diagrams including only the $C_0$ vertex as shown in Fig.~\ref{fig:twobody}.

Using the Feynman rules given above, the sum of diagrams shown
in Fig.~\ref{fig:twobody} can be expressed as
\begin{eqnarray}
\nonumber
  \mathcal{A}(E)&=&-i\, C_0\sum_{n=0}^\infty (-i C_0\, \mathcal{I}/2)^n\\
&=&\frac{-i C_0}{1+i C_0 \mathcal{I}/2}~,
\label{eq:c0geometric}
\end{eqnarray}
where $\mathcal{I}$ is given by
\begin{eqnarray}
  \mathcal{I}&=&\int\frac{\hbox{d}q_0}{2\pi}\frac{\hbox{d}q^3}{(2\pi)^3}
\frac{i}{p_0-q_0-\frac{\qvec^2}{2m}+i\epsilon}
\frac{i}{q_0-\frac{\qvec^2}{2m}+i\epsilon}\\
&=&i m\left(\frac{\Lambda}{2\pi^2}-\frac{\sqrt{-m p_0}}{4\pi}\right)~.
\label{eq:1loop}
\end{eqnarray}
This integral has been regularized with an ultraviolet cutoff
$\Lambda$.
Inserting Eq.~(\ref{eq:1loop}) in (\ref{eq:c0geometric}), we end up with the expression
\begin{equation}
  \mathcal{A}=\frac{8\pi}{m}\left[-\frac{8\pi}{m
      C_0}-\frac{2\Lambda}{\pi}+\sqrt{-m p_0}\right]^{-1}~.
\end{equation}
We reproduce the leading order term in Eq.(\ref{eq:largea}) by setting
\begin{equation}
  C_0=\frac{8\pi a}{m}\frac{1}{1-2\Lambda a/\pi}~.
\end{equation}
In the two-body system the leading order amplitude is therefore given by the
sum of all diagrams which contain only the $C_0$ vertex. Subleading
corrections can then be calculated from perturbative insertions of the higher
order operators dressed to all orders by $C_0$.

The short-range EFT power counting for the two-body sector was developed in
Refs.~\cite{vanKolck:1998bw,Kaplan:1998tg,Kaplan:1998we,Gegelia:1998gn,Birse:1999up}.
A regularization scheme alternative to cutoff regularization which makes the
powercounting on a diagram-by-diagram basis explicit was developed
in Refs.~\cite{Kaplan:1998tg,Kaplan:1998we}.

It has to be emphasized that the powercounting is of course different in the
case of a scattering length of natural size, i.e. $a\sim R$. In this case the
powercounting and therefore the order at which a particular diagram
contributes follows directly from naive dimensional analysis of the
corresponding operators. This EFT is only of limited interest in the
few-body sector, but it has found interesting applications in the many-body sector
\cite{Hammer:2000xg,Platter:2002yr}.

\subsection{The Three-Body System}
\label{subsec:3body}
Applications of the framework laid out above to the two-nucleon system have
been very successful (for a review see Ref.~\cite{Bedaque:2002mn}).

Bedaque, Hammer and Van Kolck were the first to consider the three-body system
using the short-range EFT~\cite{Bedaque:1998kg,Bedaque:1998km}. They used an
equivalent form of the Lagrange density given in Eq.~(\ref{eq:Lagrange1}) that
turns out to simplify the treatment of the three-body system. This form uses
an auxiliary field, the dimeron $T$, which has the quantum numbers of a two-body
state~\cite{Bedaque:1998kg}:
\begin{equation}
\mathcal{L}={\psi}^{\dagger}\left(i
  \partial_t+\frac{\overrightarrow{\nabla}^2}{2m}\right)\psi
+\Delta T^\dagger T-\frac{g}{\sqrt{2}}(T^\dagger\psi\psi+\hbox{h.c.})
+h T^\dagger T\psi^\dagger\psi \ldots .
\label{eq:lagrangian2}
\end{equation}
The Lagrange density above is equivalent to the density in
Eq.~(\ref{eq:Lagrange1}) if the low-energy constants are chosen to be
$2 g^2/\Delta=C_0$ and $-18 hg^2/\Delta^2=D_0$.
\begin{figure}[t]
\centerline{\includegraphics*[width=11cm,angle=0]{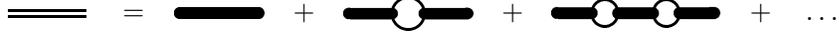}}
\caption{\label{fig:dimeron}The dressed dimeron propagator}
\end{figure}
The dimeron field $T$ has the bare propagator
\begin{equation}
  \frac{i}{\Delta}~.
\end{equation}
It is clear that the power counting introduced in the previous section dictates
a resummation of the particle-particle loops as shown in
Fig.~\ref{fig:dimeron}.  We therefore obtain the {\it dressed} dimeron
propagator~\footnote{No cutoff dependence appears in the expressions for the
  two-body propagator in Eq.~(\ref{eq:prop-dimeron}) since we assume that the
  bubble loop integrals from the two-loop sector were evaluated using
  dimensional regularization with minimal subtraction.}:
\begin{eqnarray}
  \label{eq:prop-dimeron}
\nonumber
  i D(p_0,\pvec)&=&
\frac{-i}{-\Delta+\frac{m g^2}{4\pi}\sqrt{-m
    p_0+\frac{\pvec^2}{4}-i\epsilon}}\\
&=&-\frac{4\pi}{m g^2}\frac{i}{-1/a+\sqrt{-mp_0+\pvec^2-i\epsilon}}~.
\end{eqnarray}
This propagator has a pole at $p_0=p^2/(4 m)-(4\pi \Delta/(m g^2))^2/m$.
Evaluating the residue of Eq.~(\ref{eq:prop-dimeron}) leads to the
wavefunction renormalization factor
\begin{equation}
  \label{eq:Zdimeron}
Z_D=\frac{8 \pi}{m^2 g^2 a} ~. 
\end{equation}

\begin{figure}[t]
\centerline{\includegraphics*[width=5.in,angle=0]{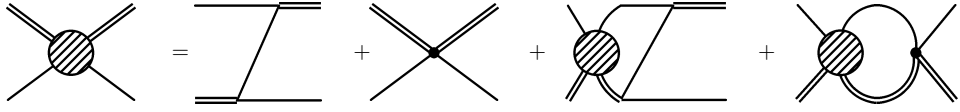}}
  \caption{\label{fig:3bodyinteq}
The integral equation for the atom-dimer scattering amplitude $t$
}
\end{figure}

Let us consider now elastic scattering between an atom and a dimer shown
diagrammatically in Fig~\ref{fig:3bodyinteq}.  Using the
Feynman rules discussed previously with the added information on the dimeron
propagator, we can consider the amplitude for elastic particle-dimer
scattering.
\begin{eqnarray}
  \label{eq:STM-off-shell}
\nonumber
   t({\bf p},{\bf k};E)&=&\frac{2m g^2}{k^2+p^2+m E+{\bf p}\cdot {\bf k}}
+h \\
\nonumber
&&\hspace{-3cm}+8\pi \int\frac{\hbox{d}^3 q}{(2\pi)^3}\frac{1}{-1/a+\sqrt{3 q^2/4-mE-i\epsilon}}
\left[\frac{t({\bf q},{\bf k};E)}{-m E+q^2+p^2+{\bf p}\cdot {\bf q}}
+\frac{h}{2mg^2}\right]~.\\
\end{eqnarray}
The reason why we have included the three-body interaction $h$ will be
discussed below.

In order to relate the amplitude to observables, we have to multiply $t$ with
the wavefunction renormalization $Z_D$ factor from Eq.~(\ref{eq:Zdimeron}). It is
therefore convenient to rewrite the integral equation:
\begin{eqnarray}
  \label{eq:3bodyscattering}
\nonumber
  t_R({\bf p},{\bf k};E)&=&\frac{16\pi}{m a}\biggl[\frac{1}{k^2+p^2-mE+{\bf p}\cdot {\bf k}}
+\frac{H(\Lambda)}{\Lambda^2}\Biggr] \\
\nonumber
&&\hspace{-2cm}+8\pi \int\frac{\hbox{d}^3 q}{(2\pi)^3}\frac{1}{-1/a+\sqrt{3 q^2/4-mE-i\epsilon}}
\left[\frac{t_R(\qvec,\kvec;E)}{-mE+q^2+p^2+{\bf p}\cdot {\bf q}}
+\frac{H(\Lambda)}{\Lambda^2}\right]~,\\
\end{eqnarray}
where $t_R=Z_D t$ and $h=2 m g^2 H(\Lambda)/\Lambda^2$.
The total energy in this process is given by $E=\frac{3}{4 m}\kvec^2-E_D$ and
the on-shell point is given by $\kvec=\pvec$.  It is straightforward to
decompose the amplitude into contributions from channels with different
orbital angular momentum quantum number $l$:
\begin{equation}
\label{eq:STM-J}
  t_l(p,k;E)=\frac{1}{2}\int_{-1}^{1}\hbox{d}x\;P_l(x)\,t(\pvec,\kvec;E)~.
\end{equation}
where $x=\pvec\cdot\kvec/(pk)$ and $P_l(x)$ denotes a Legendre
polynomial. Projecting onto $S$-waves ($l=0$) gives
\begin{eqnarray}
  \label{eq:3bdytmatrixswave}
\nonumber
  t_0(p,k;E)&=&\frac{8\pi}{m a}\biggl[\frac{1}{p k}\ln\left(\frac{p^2+p k+k^2-m E}{p^2-p
      k+k^2-m E}\right)+\frac{2 H(\Lambda)}{\Lambda^2}\biggr]\\
\nonumber
&&\hspace{1cm}+\frac{2}{\pi}\int_0^{\Lambda}\hbox{d}q\, q^2
\frac{t(q,k;E)}{-1/a+\sqrt{3 q^2/4-m E-i\epsilon}}\\
&&\hspace{1.5cm}\times\left[\frac{1}{p q}\ln\left(
\frac{p^2+p q+q^2-mE}{p^2-p q+q^2-mE}\right)
+\frac{2 H(\Lambda)}{\Lambda^2}\right]~.
\end{eqnarray}
Here a cutoff $\Lambda$ has been introduced to make the integral equation
well-defined. Equation~(\ref{eq:3bdytmatrixswave}) is then related to the
atom-dimer phase shift via
\begin{equation}
  t_0(k,k)=\frac{3\pi}{m}\frac{1}{k\cot\delta_{AD}-ik}~.
\end{equation}
Equation (\ref{eq:3bodyscattering}) (without the three-body
force) is also known as the Skorniakov-Ter-Martirosian (STM) equation, named
after the first ones to derive an integral equation interacting through
zero-range two-body interactions~\cite{STM57}.

Bound states can be found by solving the homogeneous version of the integral
equation in Eq.~(\ref{eq:3bdytmatrixswave}).
\begin{figure}[t]
\centerline{\includegraphics*[width=4.in,angle=0]{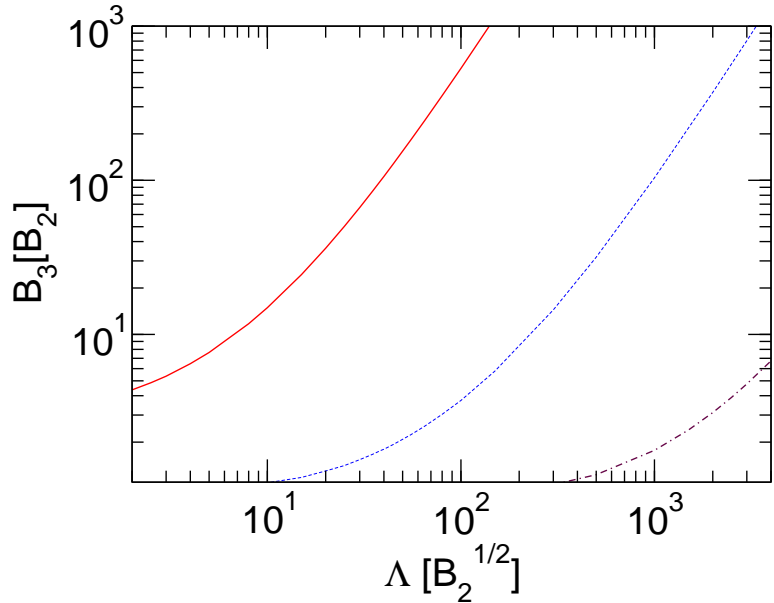}}
  \caption{\label{fig:lambdab3}
The shallowest three-body binding energies indicated by
the solid, dashed, and dash-dotted lines as a function of the 
momentum cutoff $\Lambda$. 
}
\end{figure}
So far we have not explained why we included the three-body force in our {\it
  leading order} equation. Naive dimensional analysis by counting the mass
dimension of the three-body operator suggests that the three-body force is a
higher order effect.

Let us therefore consider the bound state problem and set the three-body force
to zero. The resulting strong cutoff-dependence of three-body binding energies
is shown in Fig.~\ref{fig:lambdab3}. This cutoff-dependence is no residual
dependence which becomes weaker with increasing cutoff but is a genuine result
of the use of zero-range interactions in the three-body sector. It was in fact
already pointed out by Danilov in the 1960's that the STM equation has no unique
solution~\cite{Dan61} and it was later suggested by Kharchenko to fix the
cutoff and to treat it as a parameter~\cite{Khar73}.  The field-theoretic
perspective implies that three-body observables are sensitive to
short-distance effects that have not been properly renormalized.

Bedaque, Hammer and van Kolck \cite{Bedaque:1998kg,Bedaque:1998km} showed that the inclusion of the three-body force as in
Eq.~(\ref{eq:3bdytmatrixswave}) results in fully renormalized
observables\footnote{An alternative perspective can be won by an analysis of
  this problem in coordinate space. An analysis in the hyperspherical
  formalism shows that the three-body problem with infinite scattering length
  becomes a Schr\"odinger like problem with a $1/\mathcal{R}^2$ potential,
  where $\mathcal{R}$ denotes the three-body hyperradius. The solution to this
  problem requires a short-distance boundary condition which is determined by
  a three-body datum. An excellent discussion of the analysis of this problem
  in the hyperspherical formalism is given in \cite{NFJG01}.}.
The three-body force $H(\Lambda)$ is therefore included and its value
is determined by adjusting Eq.~(\ref{eq:3bdytmatrixswave}) to a given
three-body datum (such as the binding energy of a three-body bound state).
It was shown furthermore that the cutoff dependence of $H(\Lambda)$ can be
approximated with
\begin{equation}
  \label{eq:Hrunning}
  H(\Lambda)=-\frac{\sin\left(s_0\ln(\Lambda/L_3)-\arctan(1/s_0)\right)}
{\sin\left(s_0\ln(\Lambda/L_3)+\arctan(1/s_0)\right)}~,
\end{equation}
where $L_3$ is a parameter fixed from experiment and $s_0\approx
1.00624$. It can easily be seen that $H(\Lambda)$ is periodic in the cutoff. A
rescaling of the cutoff $\Lambda$ by a factor of $\exp(n \pi/s_0)\approx 22.7$ gives back
the same result for $H$
\begin{equation}
  H(\Lambda)=H(\Lambda \exp(n\pi/s_0))~.
\end{equation}

\begin{figure}[t]
\centerline{\includegraphics*[width=10cm,angle=0]{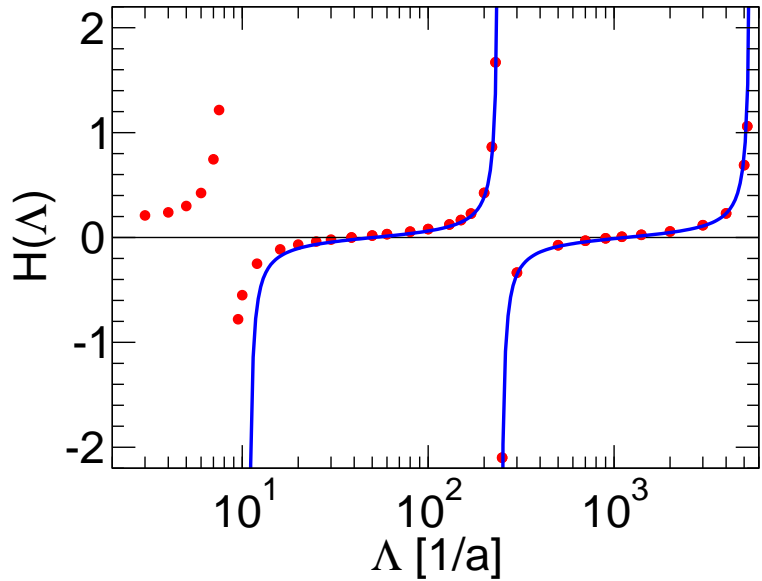}}
  \caption{\label{fig:limitcycle} Coupling constant $H$ as a function of the
    cutoff $\Lambda$. The solid shows a fit to Eq.~(\ref{eq:Hrunning}) for an
    arbitrarily chosen three-body renormalization point.}
\end{figure}
This particular running of the coupling constant $H$ is called a limit cycle
and was first disussed as an additional type of renormalization group flow in
Ref.~\cite{Wilson:1970ag}. The limit cycle is also reflected in
observables: whenever the cutoff is increased by a factor of 22.7, the number
of bound states in the spectrum increases by one (as can be seen in
Fig.~\ref{fig:lambdab3}).  In the limit of infinite scattering length, one
finds an infinite tower of three-body bound states:
\begin{eqnarray}
E^{(n)}_T = (e^{-2\pi/s_0})^{n-n_*}  \kappa^2_* /m,
\label{kappa-star}
\end{eqnarray}
where $\kappa_*$ is the binding wavenumber of the Efimov trimer labeled by
$n_*$ and $m$ is the mass of the particles. The short-range EFT reproduces
therefore at leading order the Efimov effect. Vitaly Efimov discovered in the
1970s that the zero-range limit of the 3-body problem for nonrelativistic
particles with short-range interactions shows discrete scale invariance.
If $a = \pm \infty$, there are infinitely many 3-body bound states with an
accumulation point at the 3-atom scattering threshold. These {\it Efimov
  states} or {\it Efimov trimers} have a geometric spectrum \cite{Efimov70}.
Furthermore, he pointed out that these results were also valid for finite
scattering length as long as $a\gg r_s$. The short-range EFT is therefore an
approach in which the properties of few-body systems related to Efimov physics
can be analyzed but that furthermore facilitates the systematic inclusion of
perturbations due to the finite range of the underlying interaction.

In the discussion above, we have considered systems with identical mass which
we will consider for the main part of this work. Nonetheless, it should be
pointed out that the consequences of different masses on the discrete scale
invariance/limit cycle are well-understood \cite{Braaten:2004rn}. 
\paragraph{\it One-parameter correlations:\\}
The necessity of two counterterms at leading order to obtain cutoff independent
results implies that two very different types of one-parameter correlations
exist. One can for example consider correlations between two- and three-body
observables by changing the two-body counterterm and therefore
effectively changing the scattering length. 
\begin{figure}[t]
\bigskip
\centerline{\includegraphics*[width=8cm,angle=0]{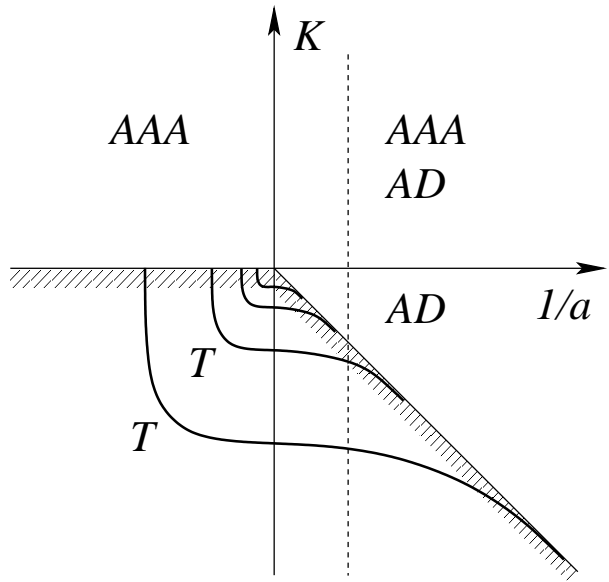}}
\medskip
\caption
{The $a^{-1}$--$K$ plane for the three-body problem. The allowed regions for
three-atom scattering states and atom-dimer scattering states are
labeled $AAA$ and $AD$, respectively. The heavy lines labeled $T$
are two of the infinitely many branches of Efimov states.
The cross-hatching indicates the threshold for scattering states.
}
\label{fig:efimovplot}
\end{figure}
A well-known example of such a correlation is the so-called Efimov plot that
is shown in Fig.~\ref{fig:efimovplot}. In this plot the binding momentum of
the $n$th trimer state $K_n=\sqrt{m E_T^{(n)}}$ is plotted against the inverse
scattering length. At the origin of the plot the three-body system shows the
afore-mentioned exact discrete scaling symmetry with respect to the binding
momenta along the y-axis. A change of $L_3$ in Eq.~(\ref{eq:Hrunning})
corresponds therefore to a rescaling of the Efimov plot.
Any point in this plot is therefore equally well-suited to characterize the
short-distance behavior of the three-body amplitude and can
be used instead of $L_3$ as the three-body parameter.
Several distinct points are particularly convenient:
\begin{itemize}
\item $\kappa_*$, the binding momenta of a trimer in the limit
  $a\rightarrow \infty$,
\item $a_*'$, the scattering length at which an Efimov trimer crosses the
  three-atom threshold,
\item $a_*$, the scattering length at which an Efimov trimer crosses the
  dimer-atom threshold~.
\end{itemize}
Other three-body parameters can of course be defined and generally all of them
are related by simple expressions to each other, e.g. $a_* \approx 0.0798
\,\kappa_*^{-1}$~\cite{Braaten:2004rn}.

A very different type of correlation plot can be generated by keeping the
two-body scattering length constant while varying the three-body force.  This
allows the study of the correlation between different three-body observables.  One
such correlation that is very well known is the Phillips line (which will be
shown later). It is an approximately linear correlation between the
particle-dimer scattering length and the three-body binding energy. This
correlation line is well-known in nuclear physics, since calculations of the
neutron-deuteron scattering length and triton binding energy with different
two-nucleon potentials lie near this line.

Correlation lines calculated using the short-range EFT will in fact
provide constraints on calculations with realistic potentials that give a large
two-body scattering length. A three-body calculation of particle-dimer
scattering length and binding energy employing such a potential has to give a
point lying on or close to the Phillips line. The short-range EFT provides
therefore strong low-energy constraints even in the absence of an experimental
datum to fix the three-body low-energy constant.

\paragraph{\it Higher Orders and Higher Partial Waves:\\}
The leading order (LO) calculations described above give results for the limit
in which the range is taken to 0.
The analysis of higher order corrections is important for several reasons.
It allows to increase the accuracy of predictions for observables but
provides also further information on the convergence radius of the
short-range EFT. It was also mentioned before that the Lagrange density
shown in Eq.~(\ref{eq:Lagrange1}) contains an infinite number of
counterterms. It is therefore natural to ask at what order the next
three-body force enters.

The calculation of higher order corrections has been addressed a number of
times. The correction to observables linear in the effective range was already
considered by Efimov using the hyperspherical formalism \cite{Efimov:1991zz,Efimov:1993zz}
but no explicit results were given for systems in which the Efimov effect is
relevant. Hammer and Mehen \cite{Hammer:2001gh} calculated the next-to-leading
order (NLO) correction
to neutron-deuteron scattering perturbatively. They found an improvement in
the description of the corresponding phase shift but didn't present a result
for the three-body binding energy at NLO. The extension of their perturbative
approach to higher orders is involved since it requires the knowledge of the
full off-shell three-body scattering amplitude.

This and and other publications displayed an unexpected convergence
pattern since the NLO shift to LO observables was unexpectedly small. This
lead to a more detailed analysis of the NLO correction to observables which is
linear in the effective range \cite{Platter:2008cx}. It was found that the
exact discrete scale invariance of the leading order wave function in the
unitary limit protects the LO bound state spectrum. That means, if
$E^{(n)}_{T,{\rm LO}}$ ($E_{T,{\rm NLO}}^{(n)}$) denotes the binding energy of the
$n$th three-body bound state at LO (NLO) in the limit $a\rightarrow \infty$, then we have
\begin{equation}
 E^{(n)}_{T,{\rm LO}}- E_{T,{\rm NLO}}^{(n)}=0 \quad \hbox{for all}\:\: n\:\: {\rm
   and}\:\: \gamma=0~.
\end{equation}
This observation was made independently by Th\o gersen, Fedorov and
Jensen~\cite{thogerson2008} using a numerical analysis of the Efimov bound state
spectrum for finite range potentials.

For arbitrary scattering length, we can define the linear shift in the binding
energy of the $n$th three-body bound state in the following way:
\begin{equation}
E_T^{(n)}=E_T^{(n*)}
\left[F_n\left(\frac{\gamma}{\kappa_*}\right)
+ \kappa_* \,r_s \,G_n\left(\frac{\gamma}{\kappa_*}\right) + {\cal O}[(\kappa\, r)^2]\right],
\label{eq:Gn1}
\end{equation}
and we have $G_n(0)=0$ for $\gamma=0$. In Ref.~\cite{Platter:2008cx} the
function $G_n$ was also analyzed for finite $\gamma$ and evidence was found
that $G_n$ is close to 0 at the atom-dimer threshold. An ongoing analysis will
shine more light on the properties of the linear range correction and will
address the impact of finite range corrections on universal relations that
relate the recombination maxima at negative scattering length to the
recombination minima at positive scattering length and will give a better
understanding of how different signatures of Efimov physics are impacted by
finite range effects \cite{Chen2}.

Bedaque {\it et al.} \cite{Bedaque:2002yg} calculated the neutron-deuteron
phase shifts up to next-to-next-to-leading order (N2LO) and included (based on
a perturbative power counting
argument that assumes the cutoff to be $\Lambda \sim 1/R$) an additional
energy-dependent three-body force at this order.  They also suggested to
include effective range corrections by modifying the two-body propagator in
the STM equation.  The two-body propagator with the effective range summed up
to all orders is given by
\begin{equation}
\label{eq:2prop-range1}
 D(p_0,\pvec)=\frac{1}{-\gamma+\frac{r_s}{2}(\gamma^2+m\, p_0-\pvec^2/4)+\sqrt{-m\, p_0 +\pvec^2/4}}~.
\end{equation}
It is easy to show that the denominator in Eq.~(\ref{eq:2prop-range1}) has two
poles. In the case of positive $\gamma$ and $r_s$ this propagator has the bound state
pole of the shallow dimer and a pole with binding energy
\begin{equation}
  \label{eq:Bspurious}
  E_{\rm spurious}=\frac{16}{m r_s^2}\left(1-\frac{1}{2}\gamma r_s+\frac{1}{16}(r_s\gamma)^2\right)~.
\end{equation}
This pole corresponds to a spurious bound state which is unphysical since in
the two-body system the short-range EFT cannot make predictions for energy
scales $\gg 1/(m R^2)$.  The spurious bound state pole in
Eq.~(\ref{eq:2prop-range1}) leads to an ill-behaved kernel when inserted into
the STM equation and the description of observables fails
\cite{Gabbiani:2001yh}.  It was therefore suggested by Bedaque {\it et al.} to
expand Eq.~(\ref{eq:2prop-range1}) in powers of the effective range:
\begin{eqnarray}
\nonumber
  \label{eq:prop-range2}
    D(p_0,\pvec)&=&\frac{1}{-\gamma+\sqrt{-m\, p_0 +\pvec^2/4}}
-\frac{r_s}{2}\frac{\gamma^2+m p_0-\pvec^2/4}{\left(-\gamma+\sqrt{-m\, p_0 +\pvec^2/4}\right)^2}
\\
&&\hspace{2.cm}
+\left(\frac{r_s}{2}\right)^2\frac{(\gamma^2+m p_0-\pvec^2/4)^2}{\left(-\gamma+\sqrt{-m\, p_0 +\pvec^2/4}\right)^3}
+\ldots~.
\end{eqnarray}
This amounts to a partial resummation of effective range corrections when
inserted into the STM equation. The difference in results for observables
between a purely perturbative calculation of effective range corrections and
this approach is expected to be of higher order.  The solution of a modified
STM integral equation leads directly to the desired observables and the
amount of effort to calculate higher order corrections does therefore not
increase compared to the solution of the leading order STM equation.

In Ref.~\cite{Platter:2006ev} a renormalization group analysis was performed
for large cutoffs in the STM equation. It lead to the conclusion that for
large cutoffs no additional three-body counterterm has to be included until
N3LO in the EFT expansion. Subsequent publications employing these results up
N2LO found good agreement with experimental measurements and calculations
using {\it realistic} interactions~\cite{Platter:2006ad,Hammer:2007kq}.

There is an obvious disagreement between the predictions for the order at
which the next three-body counterterm enters that requires a comment. In
Ref.~\cite{Bedaque:2002yg} it was argued that it enters at N2LO. In
Ref.~\cite{Platter:2006ev} the RG analysis lead to the conclusion that N3LO is
the order at which an energy-dependent three-body force has to be
included. The difference between the two analyses was the assumed size of the
cutoff, namely either $\Lambda \sim 1/r_s$ or $\Lambda \gg 1/r_s$ which can in
fact be the reason for the different conclusions. A future publication
\cite{Chen2} will address this issue and will try to reconcile the results of
Refs.~\cite{Bedaque:2002yg} and \cite{Platter:2006ev}.

An analysis of the power counting in higher partial waves was performed by
Grie\ss hammer \cite{Griesshammer:2005ga}. He defined the asymptotic exponent
$s_l$ which describes the momentum behavior of the half off-shell amplitude
for angular momentum $l$ and for large off-shell momenta $p\gg k$
\begin{equation}
  t_l(k,p)\propto k^l p^{-s_l-1}~.
\end{equation}
Using an argument based on the analysis of the perturbative evaluation of
higher order corrections of the three-body amplitude, he identified the
superficial degree of divergence of a given contribution at order $n$ to be
$n-2s_l$. A contribution at order $n$ will therefore diverge if  
\begin{equation}
\label{eq:sl}
  {\rm Re}[n-2s_l]\geq 0~.
\end{equation}
Naively, one expects the large momentum behaviour of this amplitude to be
determined by the driving term $K_l(k,p;E)$ of the STM equation
\begin{equation}
\label{eq:simplistic}
   t_l(k,p)\propto \lim_{k\rightarrow 0}K_l(k,p,3 k^2/4) \propto k^l/ p^{l+2}~,
\end{equation}
which would imply the simplistic estimate $s_{l}=l+1$.

Grie\ss hammer derived, however, an exact algrebraic equation for the asymptotic
exponent $s_l$ using a Mellin transformation of the integral equation defined
in Eq.~(\ref{eq:STM-J}) at zero energy
\begin{eqnarray}
\nonumber
  \label{eq:s}
  1=\left(-1\right)^l\;\frac{2^{1-l}}{\sqrt{3\pi}}\;
  \frac{\Gamma\left[\frac{l+s+1}{2}\right]\Gamma\left[\frac{l-s+1}{2}\right]}
  {\Gamma\left[\frac{2l+3}{2}\right]}\;
  {}_2F_1\left[\frac{l+s+1}{2},\frac{l-s+1}{2};
    \frac{2l+3}{2};\frac{1}{4}\right]\;.\\
\end{eqnarray}  
It depends only on the relative angular momentum $l$ and the function
${}_2F_1[a,b;c;x]$ is the hyper-geometric series. The order $n$ at which the
first three-body force in particular partial wave channel enters can then be
obtained from a solution of Eq.~(\ref{eq:s}) and does for a number of cases
not agree with the simplistic estimate derived from Eq.~(\ref{eq:simplistic}).

Equations (\ref{eq:s}) and (\ref{eq:sl}) hold for spinless bosons. In
Ref.~\cite{Griesshammer:2005ga} it was laid out how these formulas are
generalized to systems of spin-1/2 nucleons.
\subsection{The Four-Body System}
\label{subsec:4atoms}
It is natural to ask whether a new counterterm has to be included for every
new particle added to the problem. The question whether a four-body parameter
is required for consistent renormalization in the four-body system is
therefore a logical extension of the effort of applying the short-range EFT to
few-body systems. This pertinent question can also be paraphrased in more
general terms: do zero-range two-body interactions and one three-body
parameter lead to unique predictions for observables in the four-body sector?
\begin{table}[t]
\begin{center}
\begin{tabular}{|c||c|c||c|c|}
\hline
system & $B^{(0)}$ [mK] & $B^{(1)}$ [mK] & $B_{\rm BG}^{(0)}$ [mK] 
& $B_{\rm BG}^{(1)}$ [mK] \\ \hline\hline
$^4$He$_3$ & 127 & [2.186]   & 125.5  & 2.186\\
$^4$He$_4$ & 492 & 128  & 559.7  & 132.7\\ \hline
\end{tabular}
\end{center}
\caption{\label{tab:results}Binding energies of the $^4$He
  trimer and tetramer in mK.
  The two right columns show the Monte Carlo results by Blume and Greene 
  \cite{Blume:2000} (denoted by the index BG)
  while the two left columns show the EFT results of Platter, Hammer and Mei\ss
  ner \cite{Platter:2004qn}. The number in brackets was
  used as input to fix $L_3$.}
\end{table}

This question was addressed in the framework of the short-range EFT by
Platter, Hammer and Mei\ss ner in Ref.~\cite{Platter:2004qn}. In this work the
effective two- and three-body potentials at LO were generated and used
together with the quantum mechanical few-body equations \cite{Gloeckle:1983} to solve for the
binding energies of the four-body system. The leading order effective two-body
potential is given by
\begin{equation}
  \langle \kvec |V|\kvec'\rangle= \lambda_2 g(k)g(k')~,
\end{equation}
here $g(k)=\exp(-k^2/\Lambda^2)$ is a regulator function.  Since this is a
separable potential, the two-body problem can be solved exactly~\cite{Ziegelmann}
\begin{equation}
  t(E)=|g\rangle \tau(E)\langle g|~,
\end{equation}
with the two-body propagator $\tau$
\begin{equation}
  \tau(E)=\left[1/\lambda_2-\langle g| G_0(E)| g\rangle\right]^{-1}~,
\end{equation}
and $G_0$ denotes the three-body propagator. Observables in the two-body
sector will depend on the coupling strength $\lambda_2$ and the cutoff
$\Lambda$. For a given cutoff $\Lambda$ the coupling constant $\lambda_2$ is
then renormalized by fixing the pole position of the two-body propagator $\tau$
which amounts to fixing the binding energy of the two-body bound state.
The effective three-body potential is 
\begin{equation}
  V_3=\lambda_3|\xi\rangle\langle \xi|~,
\end{equation}
where $\lambda_3$ denotes the three-body coupling constant that has to be
adjusted to a three-body datum (such as the binding energy of the shallowest
three-body bound state) and $\langle u_1
u_2|\xi\rangle=\exp(-(u_1^2-3u_2^2/4)/\Lambda^2)$ is a regulator function
($u_1$ and $u_2$ denote here the canonical Jacobi momenta in the three-body
system \cite{Gloeckle:1983}).

Information on the necessity of a four-body parameter can be gained by
studying the regulator dependence of four-body observables.  A change in the
regulator $\Lambda$ corresponds to a modification of the short-distance
physics while the renormalization conditions on the coupling constants
$\lambda_2$ and $\lambda_3$ guarantee that low-energy observables in the two-
and three-body sector remain unchanged.  A detailed analysis shows then that
no four-body parameter is required since the values of the tetrameter binding
energies converge to well-defined values with increasing regulator $\Lambda$
\cite{Platter:2004qn}.

This approach was also used to address the four-body problem of $^4$He atoms.
$^4$He atoms have a scattering length 10 times larger than the range of the
He-He interaction. The existence of $n$-body clusters of $^4$He atoms was
shown experimentally \cite{STo96}, however, the measurement of the binding
energies of these cluster states is currently not possible. Several potentials
that are believed to describe the two-body interaction accurately have been
developed and have been used in Faddeev~\cite{RoY00,Ro00,MSSK01,Kolganova01}
and Monte-Carlo~\cite{Blume:2000} calculations.  The four-body binding
energies were computed, using the results for the dimer and trimer binding
binding energies obtained using the LM2M2 potential in Ref.~\cite{Blume:2000}
as input. For the three-body coupling constant, the excited three-body state of
the $^4$He trimer was chosen as input parameter.  The results can be seen in
Table \ref{tab:results}. The binding energies of the $^4$He tetramer were
found to be in good agreement with the results for the ground and excited
state of the $^4$He tetramer obtained by Blume and Greene \cite{Blume:2000}.
\begin{figure}[t]
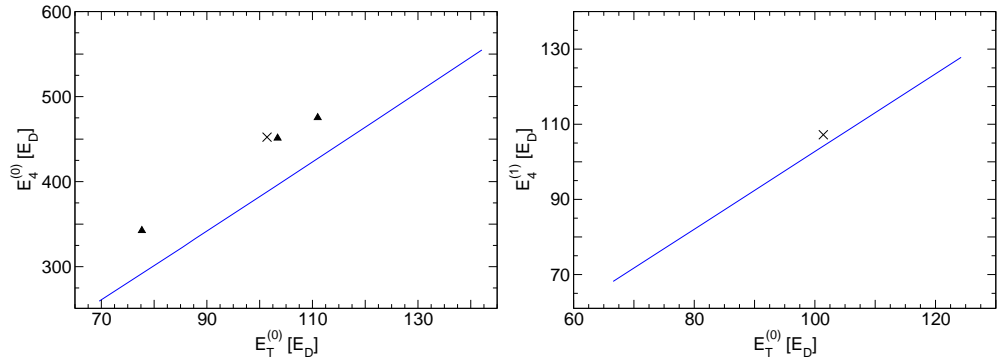

\centerline{
\includegraphics*[width=6.5cm,angle=0]{tjon_40vs30.eps}
\includegraphics*[width=6.5cm,angle=0]{tjon_41vs30.eps}}
  \caption{\label{fig:tjon-bosons}
One-parameter correlations of three- and four-boson bound states.
The crosses are results for LM2M2 potential~\cite{Blume:2000}. The triangles
are the results for the TTY, HFD-B and HFDHE2 potentials \cite{Lew97,Naka83}.
}
\end{figure}

%
By keeping the two-body parameter fixed and varying the three-body coupling
constant, the correlation line between the tetramer and trimer binding energies
can be generated. These are are approximately linear correlations which are
well known from calculations in the few-nucleon sector. They parameterize the
common knowledge that an internucleon potential that gives the correct value
for the triton binding energy also gives a good result for the four-nucleon
bound state (i.e. the $\alpha$-particle). In Fig.~\ref{fig:tjon-bosons} we
show this correlation line (which in nuclear physics is called the Tjon line)
generated for values of the three-body binding energy close to the value of
$^4$He trimer binding energy. The Tjon line is therefore a common feature of
systems with a large two-body scattering length and does not depend on any
details of the interaction at short-distances.

This approach was used furthermore for a more detailed analysis of the
four-boson system with large positive and large negative scattering
length~\cite{Hammer:2006ct}.  Results in this analysis also lead to the
conclusion that every trimer state is tied to two universal tetramer states
with binding energies related to the binding energy of the next shallower
trimer:
\begin{equation}
  \label{eq:4body1}
  E_{4,0}\sim 5\, E_T\quad {\rm and} \quad  E_{4,1}\sim 1.01\, E_T
\quad{\rm for}\quad \gamma\sim 0~,
\end{equation}
where $E_{4,0}$ denotes the binding energy of the deeper of the two tetramer
states and $E_{4,1}$ the shallower of the two.

A recent calculation by von Stecher, d'Incao and Greene \cite{Stecher:2008}
supports the findings made in \cite{Platter:2004qn,Hammer:2006ct}. The authors
of this work extended previous results to higher numerical accuracy. They
furthermore considered the relation between universal three- and four-body
bound states in the exact unitary limit ($a\rightarrow\infty$). They found
 \begin{equation}
  \label{eq:4body2}
  E_{4,0}\approx 4.57 E_T\quad {\rm and} \quad  E_{4,1}\approx 1.01 E_T~,
\end{equation}
which agree with the results obtained in Ref.~\cite {Hammer:2006ct} and given
in Eq.~(\ref{eq:4body1}).

\begin{figure}[tb]
\centerline{\includegraphics*[width=10cm,angle=0]{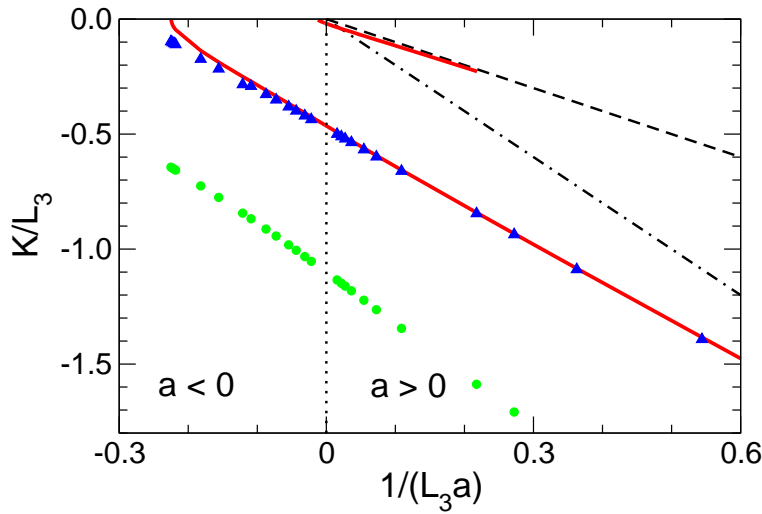}}
\caption{\label{fig:efimov-4body} The $a^{-1}$--$K$ plane for the four-body
  problem. The circles and triangles indicate the four-body ground and excited
  state energies $B_4^{(0)}$ and $B_4^{(1)}$, while the lower (upper) solid
  lines give the thresholds for decay into a ground state (excited state)
  trimer and a particle. The dash-dotted (dashed) lines give the thresholds
  for decay into two dimers (a dimer and two particles).  The vertical dotted
  line indicates infinite scattering length.  All quantities are given in
  units of the three-body parameter $L_3$.  }
\end{figure}

The results obtained by the Hammer and Platter in Ref.~\cite{Hammer:2006ct}
were furthermore presented in the form of an extended Efimov plot, shown in
Fig.~\ref{fig:efimov-4body}. Four-body states have to have a binding energy
larger than the one of the deepest trimer state. The corresponding threshold
is denoted by lower solid line in Fig.~~\ref{fig:efimov-4body}.  The threshold
for decay into the shallowest trimer state and an atom is indicated by the
upper solid line. At positive scattering length, there are also scattering
thresholds for scattering of two dimers and scattering of a dimer and two particles
indicated by the dash-dotted and dashed lines, respectively.  The vertical
dotted line denotes infinite scattering length. A similar but extended version
of this four-body Efimov plot was also presented by Stecher, d'Incao and
Greene in Ref.~\cite{Stecher:2008}.
They computed also the scattering lengths at which the binding energies of
the tetramer states become zero and found
\begin{equation}
\label{eq:tetra-scatteringlengths}
  a^*_{4,0}\approx 0.43 a_*\quad{\rm and}\quad a^*_{4,1}\approx 0.92 a_*~.
\end{equation}

A further interesting detail was pointed out in \cite{Stecher:2008}. In the
unitary limit the shallower of the two-four-body states can be considered as a
trimer state with an extra particle attached to it. If this interpretation as
an effective two-body state holds, then it leads to an important
conclusion. Such an effective two-body picture in terms of a heavy and a light
particle implies that the atom-trimer scattering length has to be large. For a
two-body system with large positive scattering a length $a$ and unequal
masses, the binding energy is $1/(2 \mu a^2)$, where $\mu=m_1 m_2/(m_1+m_2)$
is the reduced mass. With $m_1=m$ and $m_2=3m$ we obtain for  $a_{\rm AT}$
\begin{equation}
  E_T-E_{4,0}\approx\frac{2}{3}\frac{1}{m a_{\rm AT}^2}~,
\end{equation}
where $a_{aT}$ denotes the atom-trimer scattering length. We can therefore
obtain a simple estimate for the scattering length of an Efimov
trimer with index $n$ and an atom in the unitary limit
\begin{equation}
  a_{\rm AT}^{(n)}=\sqrt{\frac{2}{3}\frac{1}{m (E_T-E_{4,0})}}
\approx\sqrt{\frac{2/3}{0.01\,m E^{(n)}_T}}
\approx 8.2 \,\kappa_n^{-1}~.
\end{equation}
Since $E_T^{(n)}$ is the only scale in the problem, naive dimensional analysis
predicts the range of the atom-trimer interaction to be of order
$\kappa_n^{-1}$.  The estimated scattering length $a_{\rm AT}$ is thus by an
order of magnitude larger than this estimate of the range.
\section{Low-Energy Universality in Atomic Physics}
\label{sec:amo}
In this section we will discuss recent applications of the short-range EFT to
systems of cold atoms. We will focus in particular on the problems of
three-body recombination and atom-dimer relaxation in ultracold gases but will
also report on recent progress in the four-body sector.

The recombination rate and atom-dimer relaxation have been identified as key
signatures in the search for Efimov physics since they can be obtained my
measuring atomic loss rates. Three-body recombination is a process in which a
two-body bound state is formed as the result of a three-body collision. The
two outgoing particles will gain kinetic energy in this process equal to the
binding energy of the dimer. In experiments with ultracold atoms, the kinetic
energy is often sufficient to allow the atom and dimer to subsequently escape
the trapping potential.

Ultracold gases of alkali atoms are very well suited for such experiments
since in many cases
Feshbach resonances allow the scattering length to be tuned to
arbitrarily large values using an external magnetic field. Observables can
therefore be measured as a function of the two-body scattering length and
results can be compared to the corresponding one-parameter correlation
discussed previously.

The first experimental evidence for Efimov physics in an atomic system was
presented by Grimm and co-workers \cite{Grimm06}.  In this pioneering experiment
with ultracold $^{133}$Cs atoms in the lowest hyperfine state, they observed a
resonant enhancement in the 3-body recombination rate at $a \approx -850~a_0$
that can be explained with an Efimov trimer close to the 3-atom threshold.

Since then the number of experiments that show evidence for universal
three-body physics has increased significantly. Signatures of Efimov physics
have been found in three-component Fermi gases of
$^6$Li~\cite{Heidelberg,PennState}, in a Bose gase of $^{39}$K
atoms~\cite{efimov-K41}, and in heterogenuous mixtures of $^{41}$K and
$^{87}$Rb~\cite{efimov-mixture}. The level of sophistication in experiments
has increased so much that recent measurements even test the implications of
low-energy universality on four-body dynamics~\cite{Ferlaino:2009}.

One experimental tool that has become essential for the analysis of Efimov
physics are Feshbach resonances. As mentioned above they allow the
interparticle scattering length to be tuned by adjusting an external magnetic
field. A Feshbach resonance arises due to the coupling of two atoms in an
open channel to a closed channel. The open channel corresponds to a pair of
atoms in energetically allowed hyperfine states while the closed channel
corresponds to a combination of hyperfine states that is energetically inaccessible to
asymptotic scattering states. The coupling between the open and closed
channel arises due to the hyperfine interaction. The scattering length becomes
large when a magnetic field is used to tune a bound state in the closed
channel to the threshold of the open channel. The dependence of the scattering
length on the magnetic field near a Feshbach resonance can be described by
\begin{equation}
  \label{eq:feshbach}
  a(B)\approx a_{\rm bg}\left(1-\frac{\Delta}{B-B_0}\right)~,
\end{equation}
where $B_0$ denotes the scattering at which a state in the closed channel is
at threshold and $\Delta$ governs the width of the resonance. For magnetic fields away
from the resonance $B_0$, the scattering length is given by $a_{\rm
  bg}$. The scattering length can therefore be tuned to values significantly
larger than the low-energy length scale of the atom-atom interaction which is
given by the so-called van der Waals length $\ell_{\rm vdW}$. This quantity
is related to the long-range van der Waals tail of the atom-atom interaction via
\begin{equation}
  V(r)\:\rightarrow\:-\frac{\ell_{vdW}^4/m}{r^6}~.
\end{equation}
This length scale sets therefore the natural scale for the effective range
$r_s\sim R$ of the interaction and quantifies at which distances the
short-range EFT is not applicable anymore.  
\subsection{Three-Body Recombination of Identical Bosons}
In a three-body recombination process, three particles
collide and two of them form a two-body bound state. If the scattering length
is large and positive, the resulting two-body bound state can
be a {\it deep} dimer or a {\it shallow} dimer when the scattering length is
positive. A deep dimer is a two-body bound state that cannot be described
within the short-range EFT and has binding energy $\gtrsim 1/(m R^2)$. The
shallow dimer can be described within the short-range EFT and has binding
energy $\sim 1/(m a^2)$. If the scattering length is large and negative the
resulting two-body bound state can only be a deep dimer.

\begin{figure}[t]
\centerline{\includegraphics*[width=5.in,angle=0]{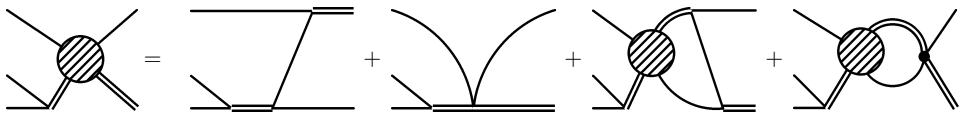}}
  \caption{\label{fig:reco-inteq}
The integral equation for three-body recombination into the shallow dimer.
}
\end{figure}

The change in density $n$ due to such losses is described by
\begin{equation}
  \label{eq:recobosons}
  \frac{\hbox{d}}{\hbox{d}t}n=-n_{\rm lost}\alpha\: n~,
\end{equation}
where $n_{\rm lost}\alpha\equiv L_3$ is an experimentally measurable loss rate constant,
$n_{\rm lost}$ is the number of atoms escaping in every recombination process
and $\alpha$ is the event rate of the recombination process. In the case of a
Boltzmann distribution, the event rate can be related to the hyperangular
averaged recombination rate $K(E)$:
\begin{equation}
\alpha(T) \approx 
\frac{\int_0^\infty dE \, E^2 \, e^{-E/(k_B T)} \, K(E)}
    {6 \int_0^\infty dE \, E^2 \, e^{-E/(k_B T)}} \,.
\label{alpha-T}  
\end{equation}
The recombination rate $K(E)$ can be decomposed into a contribution from recombination
into the shallow dimer and a contribution from recombination into deep dimers:
\begin{equation}
  K(E)=K_{\rm shallow}(E)+K_{\rm deep}(E)~.
\end{equation}
The three-body recombination rate into the shallow dimer can be decomposed
into channel contributions with different total orbital angular momentum $J$:
\begin{equation}
  K_{\rm shallow}(E)=\sum_{J=0}^\infty K^{(J)}(E)~.
\end{equation}
The recombination rate $K_{\rm shallow}$ into the shallow dimer
is then related to the total atom-dimer breakup cross section via 
\begin{equation}
\label{eq:K-cross}
  K_{\rm shallow}(E)=\frac{192\sqrt{3}\pi(E_D+E)}{m^2 E^2}\,\sigma_{\rm breakup}(E)~.
\end{equation}
This implies that the recombination rate is related to the S-matrix for the
scattering of three atoms into an atom and dimer through the following
relation:
\begin{equation}
  K^{(J)}(E) = \frac{144 \sqrt{3} \pi^2  (2J+1)}{m^3 E^2}
\sum_{n=3}^\infty \left| S_{AAA,AD}^{(J,n)}(E) \right|^2.
\label{Kshallow-S}
\end{equation}
Here, $n$ denotes a set of quantum numbers that includes the relative angular
momenta between the particles. The hyperangular average is implemented by the
sum over $n$ which starts at $n=3$ for convenience.  The unitarity of the
S-matrix in the angular momentum $J$ sector implies
\begin{equation}
|S_{AD,AD}^{(J)}(E)|^2 
+ \sum_{n=3}^\infty |S_{AD,AAA}^{(J,n)}(E)|^2 = 1 .
\label{unitarity:0}
\end{equation}
By unitary, the recombination rate into the shallow dimer is therefore directly
related to the S-matrix element for elastic atom-dimer scattering:
\begin{equation}
K^{(J)}(E) = \frac{144 \sqrt{3} \pi^2  (2J+1)}{m^3 E^2}
\left( 1 - \Big| e^{2 i \delta_{AD}^{(J)}(E)} \Big|^2 \right) ,
\label{K-deltaAD}
\end{equation}
since
\begin{equation}
S_{AD,AD}^{(J)}(E) = e^{2 i \delta_{AD}^{(J)}(E) } .
\label{S-deltaAD}
\end{equation}
The phase shifts for atom-dimer scattering are therefore sufficient for a
calculation of the recombination rate into shallow dimers.
\paragraph{\it Recombination into the Shallow Dimer:\\}
Bedaque, Braaten and Hammer were the first to use the short-range EFT to describe
three-body recombination into the shallow dimer \cite{Bedaque:2000ft}.  They
calculated the three-body recombination rate at zero temperature and for $J=0$
by solving the integral equation
for the scattering of three atoms into an atom and dimer,
shown diagrammatically in Fig.~\ref{fig:reco-inteq}.
It can be shown that for zero energy, the amplitude for this process is
related to the amplitude for elastic scattering given in
Eq.~(\ref{eq:3bdytmatrixswave}) at an off-shell point. This can be used to
simplify the numerical calculation of the three-body recombination
rate. This approach can also be used to calculate the contribution to the
recombination rate from channels with $J\neq 0$. The contribution to the
three-body recombination rate is in this case
\begin{equation}
K^{(J)}(E) = \frac{144 \sqrt{3} \pi^2 (2J+1) f_J(x)}{x^4} \, 
\frac{ a^4}{m} \,,
\label{alpha-J:uni}
\end{equation}
where $f_J(x)$ is a real-valued scaling function:
\begin{equation}
f_J(x) = 1 - \exp \big(-4 \, {\rm Im} \, \delta_{AD}^{(J)}(E) \big)  \, ,
\label{fJ-delta}
\end{equation}
and $x$ is defined as
\begin{equation}
x = \left( m a^2 E \right)^{1/2} .
\label{x-def}
\end{equation}
As $x \to 0$, the leading powers of $x$ are determined by Wigner's
threshold law \cite{EGS01}: $f_J(x) \sim x^{2 \lambda_J+4}$, where $\lambda_1
= 3$ and $\lambda_J = J$ for $J \ge 2$.

It is only in the channel with total angular momentum $J=0$ that observables
depend on the three-body parameter. In this channel, a different
perspective on the recombination rate can be gained using Efimov's radial
\cite{Efimov79,Braaten:2004rn} laws which expresses the S-matrix in the
three-body sector as a combination of universal functions of a scaling
variable $x$ and a phase $\theta_{*0}$:
\begin{subequations}
\begin{eqnarray}
S_{AD,AD}^{(J=0)}(E) & = & s_{22}(x)
+ \frac{s_{21}(x)^2 e^{2i \theta_{*0}}}
      {1 - s_{11}(x) e^{2i \theta_{*0}}}  \,,
\label{RL:AD}
\\
S_{AD,AAA}^{(J=0,n)}(E) & = & s_{2n}(x) 
+ \frac{s_{21}(x) s_{1n}(x) e^{2i \theta_{*0}}}
      {1 - s_{11}(x) e^{2i \theta_{*0}}}  \,.
\label{RL:ADAAA}
\end{eqnarray}
\end{subequations}
The phase $\theta_{*0}$ is related to the minimum in the recombination rate
$a_{*0}$ for positive scattering length
\begin{equation}
  \theta_{*0}=s_0\ln(a/a_{*0})~,
\end{equation}
where $a_{*0}\approx 0.32 \kappa_*^{-1}$ \cite{Braaten:2004rn}.

Using analytical results on the atom-dimer phase shift at breakup
threshold obtained in Ref.~\cite{MOG05}, Efimov's radial laws can be used to
derive an analytical result for the recombination rate into the
shallow dimer
\begin{eqnarray}
\label{eq:reco-rate-analytical}
K^{(0)}(E=0) = 
\frac{768 \pi^2 (4 \pi - 3 \sqrt{3}) \sin^2[s_0 \ln(a/a_{*0})]}
    {\sinh^2(\pi s_0) + \cos^2[s_0 \ln(a/a_{*0})]} \,
\frac{a^4}{m} \,.
\label{K0-analytic}
\end{eqnarray}
Macek, Ovchnikov and Gasaneo derived this analytic result for the three-body
recombination rate at the three-atom threshold in \cite{MOG06}. This
equation was also derived independently by Petrov \cite{Petrov-octs}.
Simpler approximate expressions that correspond essentially to omitting the
$\cos^2$ term in the denominator have been previously derived by Nielsen and
Macek~\cite{NM-99}, Esry, Green and Burke~\cite{EGB-99}
and by Bedaque, Braaten and Hammer~\cite{Bedaque:2000ft}.

For a fixed scattering length, the recombination rate in
Eq.~(\ref{eq:reco-rate-analytical}) has therefore the maximum value
\begin{equation}
  \label{eq:Kmax}
  K_{\rm max}=6 C_{\rm max} a^4/m~ \quad \hbox{with}
\quad C_{\rm max}=\frac{128\pi^2(4\pi-3\sqrt{3})}{\sinh^2(\pi s_0)}~.
\end{equation}

\paragraph{\it Effects of Deep Dimers:\\}
While recombination into the shallow dimer can only occur for positive
scattering length, recombination into deeply bound states (into deep dimers)
can occur for either sign of the scattering length.  Properties of these bound
states cannot be calculated in the short-range EFT. However, their cumulative
effects on low-energy observables can be accounted for analytically continuing
the Efimov parameter to complex values. This introduces only one additional
real-valued parameter denoted usually with $\eta_*$. If the dependence of an
amplitude on $\kappa_*$ or, equivalently, $a_{*0}$ is known analytically, the
effect of deep dimers can be taken into account by the simple substitution
\begin{equation}
  \label{eq:eta1}
  \ln a_{*0}\longrightarrow \ln a_{*0}-i\eta_*/s_0
\end{equation}

For positive scattering length, making this replacement in the amplitudes for
recombination into the shallow dimer leads to
\begin{eqnarray}
\label{eq:Kshallow-apos}
K_{\rm shallow}(0) &=& 
\frac{768 \pi^2 (4 \pi - 3 \sqrt{3})
(\sin^2 [s_0 \ln(a/a_{*0})]  + \sinh^2\eta_*)}
{\sinh^2(\pi s_0 + \eta_*) + \cos^2 [s_0 \ln(a/a_{*0})]}\,
\frac{a^4}{m} \,,
\label{K-shallow:eta}
\end{eqnarray}
Using unitarity, one can also obtain an analytic expression for the
recombination rate into deep dimers:
\begin{eqnarray}
\label{eq:Kdeep-apos}
K_{\rm deep}(0) &=& 
\frac{384 \pi^2 (4 \pi - 3 \sqrt{3}) \coth(\pi s_0) \sinh(2\eta_*)}
    {\sinh^2(\pi s_0 + \eta_*) + \cos^2 [s_0 \ln(a/a_{*0})]}\,
\frac{a^4}{m}~, \quad (a>0)~.
\label{K-deep:eta}
\end{eqnarray}
\label{K:eta}
These results were first derived by Braaten and
Hammer~\cite{Braaten:2004rn}. Simpler approximate expressions that correspond
essentially to omitting the $\cos^2$ in the denominator were derived in
Ref.~\cite{Braaten:2003yc}. The weak dependence on the three-body parameter
$a_{*0}$ in Eq.~(\ref{eq:Kdeep-apos}) was first observed in a numerical
calculation for the case of infinitesimal $\eta_*$ \cite{Braaten:2001hf}.

For negative scattering length , there is no shallow dimer.
The recombination rate into deep dimers is given by
\begin{equation}
  \label{eq:K-deep-aneg}
  K_{\rm
    deep}(0)=\frac{765\sinh(2\eta_*)}{\sin^2(s_0\ln(a/a_*'))+\sin^2\eta_*}\frac{
    a^4}{m}~,\quad(a<0)~.
\end{equation}
This results was first derived by Braaten and Hammer in Ref.~\cite{Braaten:2003yc}.
The scaling of $K_{\rm deep}$ with $a^4$ was predicted by Nielsen and Maceck
and by Esry {\it et al.} \cite{NM-99,EGB-99}. Esry {\it et al.} were the first
to point out the existence of a log-periodic sequence of resonances related to
Efimov trimers \cite{EGB-99}. 
\paragraph{\it Three-Body Recombination at Finite Temperature:\\}
The experimental measurements of the three-body recombination rate of
$^{133}$Cs atoms were performed at finite temperature. Although the zero temperature
equations can be applied immediately to their results, it is desirable to
extend these to finite temperature.
\begin{figure}[t]
\centerline{\includegraphics*[width=10cm,angle=0,clip=true]{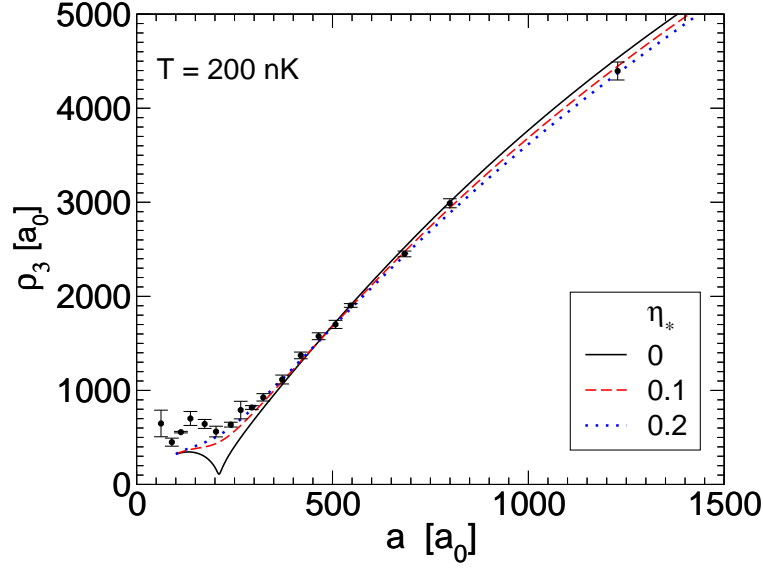}}
\vspace*{0.0cm}
\caption{The 3-body recombination length $\rho_3$ for $^{133}$Cs atoms as a
  function of $a$ for $T=200$ nK.  The data points are from
  Ref.~\cite{Grimm06}.  The curves are the universal prediction for three
  values of $\eta_*$: 0 (solid line), 0.1 (dashed line), and 0.2 (dotted
  line).\label{fig:alpha-Cs}}
\end{figure}
This requires the calculation of the energy dependent recombination rate.  For
recombination into the shallow dimer, this can be obtained by solving the
amplitude for scattering of three atoms into atom and dimer as a function of
the external momenta, calculating the thermal average of the recombination
rate, and then adjusting the three-body parameters $\kappa_*$ and $\eta_*$ to
fit experimental data. An alternative path is to determine the universal
functions $s_{ij}(x)$ from the atom-dimer scattering phase shifts.  Once these
functions are determined, the temperature dependent recombination rate can be
determined for any system of identical bosons with arbitrary three-body
parameter. This was done in Ref.~\cite{Braaten:2008kx} for the case of
positive scattering length. Using the STM equation the atom-dimer phase shifts
were calculated for a wide range of energies and three-body parameters. The
results were then fit to the formula
\begin{eqnarray}
\exp \left( 2 i \delta_{AD}^{(J=0)}(E) \right)  &=& 
s_{22}(x)  
+ \frac{s_{12}(x)^2 \exp[2is_0 \ln(a/a_{*0})]}{1 - 
  s_{11}(x) \exp[2is_0 \ln(a/a_{*0})]}
\label{expdelta-x}
\end{eqnarray}
to determine the universal scaling functions $s_{11}(x)$, $s_{12}(x)$ and
$s_{22}(x)$ for $0<x<10$.

Using the unitarity of the S-matrix it can then be shown that these three
universal scaling functions are sufficient for the calculation of the $J=0$
contributions to the recombination rate into shallow dimers, even if there are
deep dimers:
\begin{eqnarray}
K^{(0)}(E) &=& \frac{144 \sqrt{3} \pi^2}{x^4}
\biggl( 1 
- \biggl| s_{22}(x) + \frac{s_{12}(x)^2 e^{2i \theta_{*0} -2 \eta_*}}
                     {1 - s_{11}(x) e^{2i \theta_{*0} -2 \eta_*}} \biggr|^2
\\
&&\hspace{3.8cm}- \frac{(1 - e^{-4 \eta_*}) |s_{12}(x)|^2}
      {|1 - s_{11}(x) e^{2i \theta_{*0} -2 \eta_*}|^2} \biggr) 
\frac{ a^4}{m} \,,
\label{Kshallow-s}
\end{eqnarray}
They are also sufficient to calculate the recombination rate into deep dimers:
\begin{eqnarray}
K_{\rm deep}(E) = 
\frac{144 \sqrt{3} \pi^2 (1 - e^{-4 \eta_*}) 
	\big( 1 - |s_{11}(x)|^2 - |s_{12}(x)|^2 \big)} 
    {x^4 |1 - s_{11}(x) e^{2i \theta_{*0} -2 \eta_*} |^2} \, 
\frac{a^4}{m} \,.
\label{Kdeep-s}
\end{eqnarray}
The knowledge of three universal scaling functions facilitates therefore the
calculation of the temperature dependent recombination rate constant in the
presence of deep dimers.  In Fig.~\ref{fig:alpha-Cs} the theoretical results
for three different values of $\eta_*$ are compared to the experimental
results by Grimm and co-workers.  The vertical axis is the recombination
length $\rho_3$ defined by
\begin{equation}
\rho_3 = \left( \frac{2m}{\sqrt{3}} n_{\rm lost} \alpha \right)^{1/4}~,
\label{rho3}
\end{equation}
where $n_{\rm lost}$.

The recombination minimum that determines the three-body body input is near
$200~a_0$. This is comparable in size to the Waals length scale
$(mC_6)^{1/4} \approx 200~a_0$, so range corrections may be large near
the minimum and the disagreement between theoretical and experimental results
at smaller values of the scattering length is hardly surprising.
The authors of Ref.~\cite{Braaten:2008kx} were not able to determine $\eta_*$ since 
their fit was insensitive to the value of $\eta_*$. It yielded, however,
the upper bound $\eta_* < 0.2$. Their result for the recombination minimum
$a_{\rm min} = 210 (10)~a_0$ obtained from a fit for $a > 500~a_0$ agrees with
direct experimental loss measurements preformed by the Innsbruck group.
\paragraph{\it Finite Range Effects in Three-Body Recombination:\\}
The recombination rate measurements performed by Grimm and co-workers were
carried out at scattering lengths at which the finite range of the atom-atom
interaction is expected to play a significant role. The range of the
$^{133}$Cs interaction is of the order of 200~$a_0$. It is therefore of
interest to understand the impact of the finite range of the atom-atom
interaction on experimentally measurable quantities.

A first effort to calculate range corrections to the recombination rate into
shallow dimers was performed in \cite{Hammer:2006zs}. The authors considered
corrections up to N2LO in the EFT expansion to the recombination rate and
calculated the correlation between the atom-dimer scattering length and the
recombination rate $K(0)$ at zero energy. The approach was furthermore used to
calculate the recombination rate coefficient for $^4$He atoms for which the
effective range is known. This approach can be applied to more interesting
systems such as $^{133}$Cs or $^6$Li provided the corresponding effective
range is known.

\subsection{Three-Body Recombination of Fermions}
A large number of experiments with cold atoms are now carried out with
fermionic atoms.  Systems of fermionic atoms are of interest due to their
relation to solid state physics. In particular, systems with two spin states
have received considerable attention. Many-body systems of cold atoms display
superfluidity at sufficiently low temperatures. Feshbach resonances can be
used to study how the mechanism for superfluidity depends on the
interactions. In the case of fermions with two spin states, as the scattering
length is varied from $1/a <0$ to $1/a>0$ through the Feshbach resonance, the
mechanism changes continuously from the formation of Cooper pairs to the
Bose-Einstein condensation of shallow dimers. 

The Efimov can not occur in systems of fermions with only two spin states.
This can be understood from the fact that a pointlike three-body $S$-wave
interaction is forbidden due to the Pauli principle.  However, in the case of
fermions with three spin states the Efimov effect can occur again since a
pointlike three-body $S$-wave interaction is not forbidden by the Pauli
principle. Experiments using $^6$Li atoms in the three lowest hyperfine states
have been performed by Selim and co-workers \cite{Heidelberg} and O'Hara and
co-workers \cite{PennState}. The three corresponding scattering lengths in
this system have broad Feshbach resonances at nearby values of the magnetic
field, i.e. at 690~G, 810~G and 833~G. A narrow loss feature near 130~G was
discovered independently by both groups. At these magnetic field strengths all
three scattering lengths are negative and relatively large, universal results
are therefore relevant in this region and the short-range EFT is applicable.

The rate equations for the number densities $n_i$ of atoms in the three spin
states are
\begin{equation}
\frac{d\ }{dt} n_i = - K_3 n_1 n_2 n_3.
\label{dndt}
\end{equation}
In the case of this more complicated system, no analytical results
for the recombination rate are available so one has to do a
numerical calculation. The three-body recombination is related
to the amplitudes for elastic scattering 
\begin{equation}                               
K_3  = \frac{32 \pi^2}{m} \sum_{i,j} a_i a_j {\rm Im} A_{ij}(0,0),
\label{K3-A}
\end{equation}
The amplitudes $A_{ij}$ have to be calculated from a set of nine
coupled integral equations which are an extension of the STM
equation:
\begin{eqnarray}
\nonumber
\label{eq:stm-3fermions}
  A_{ij}(p,0)&=&\frac{1-\delta_{ij}}{p^2}+\frac{2}{\pi}\sum_k
  (1-\delta_{kj})\int_0^\Lambda\hbox{d}q 
\frac{q}{2 p}\ln\left(\frac{p^2+p q+q^2}{p^2+p q+q^2}\right)\\
&&\hspace{5cm}\times\frac{A(q,0)}{-1/a_k+\sqrt{3}/2 q}~.
\end{eqnarray}
\begin{figure}[t]
\centerline{\includegraphics*[height=7.0cm,angle=0,clip=true]{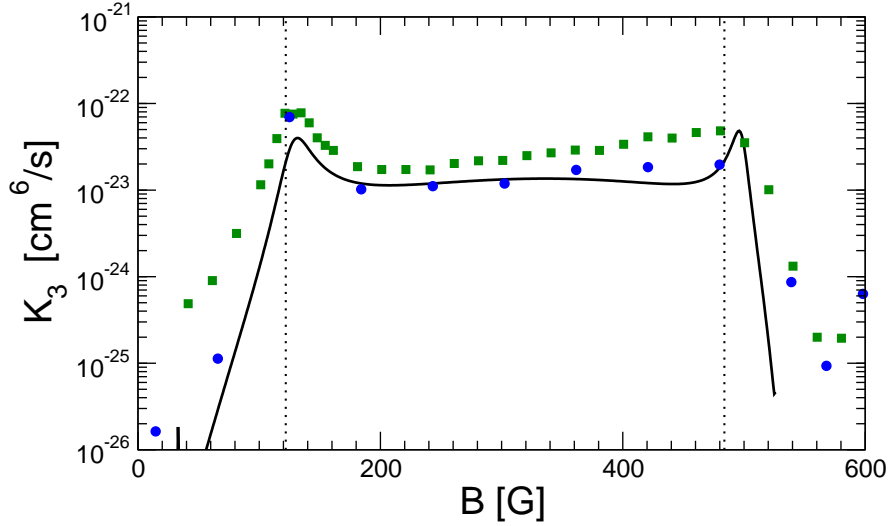}}
\vspace*{0.0cm}
\caption{ The 3-body recombination rate constant $K_3$ for the three lowest
  hyperfine spin states of $^6$Li atoms as a function of the magnetic field
  $B$.  The two vertical dotted lines mark the boundaries of the region in
  which $|a_{12}| > 2~\ell_{\rm vdW}$.  The solid squares are data points from
  Ref.~\cite{Heidelberg}.  The solid dots are data points from
  Ref.~\cite{PennState}.  The curve is a 2-parameter fit to the shape of the
  data from Ref.~\cite{Heidelberg}.}
\label{fig:Klow}
\end{figure}
These equations are written out without the use of an explicit three-body
force and instead the momentum cutoff is used to fix the the three-body coefficient.
In \cite{Braaten:2008wd} the recombination rate of general systems of three
identical fermions with three spin states was analyzed.  The case of $^6$Li
was also considered. Theoretical results on the magnetic field dependence of
the three scattering lengths were used to calculate the three-body
recombination rate in regions of the magnetic field relevant to the Penn-State
and and Heidelberg experiments. The effects of deep dimers were included by using
a complex cutoff $\Lambda e^{i\eta_*/s_0}$ in the STM integral equation
in Eq.~(\ref{eq:stm-3fermions}). The three-body parameters were adjusted to
reproduce the observed recombination maximum near 130 G.  In
Fig.~\ref{fig:Klow} we show a comparison of theoretical and experimental
results. The results are in excellent agreement with the observed
recombination rate near the narrow loss feature at 210~G. They also predict a
second narrow loss feature near 500~G. In the experiments a broad loss feature is
observed near 500~G, but the behavior in this region is not correctly
reproduced by theory. The reason for this discrepancy is under investigation.

Following this work, the problem was also considered using a wave function
approach \cite{Naidon2008} and functional renormalization
\cite{Floerchinger2008}.  The authors of both studies found qualitative
agreement with the EFT results.
\subsection{Atom-Dimer Relaxation}
Another observable closely related to three-body recombination is atom-dimer
relaxation. If there is an Efimov trimer close to the three-atom threshold,
inelastic scattering processes of atom-dimer to atom-(deep dimer) will be
resonantly enhanced. The parameters that determine the rate of these losses
are $a$, $\kappa_*$ and $\eta_*$.

The relaxation rate event constant is defined by the following equation:
\begin{equation}
  \frac{\hbox{d}}{\hbox{d}t}n_A=\frac{\hbox{d}}{\hbox{d}t}n_D=-\beta n_A n_D~,
\end{equation}
where $n_A$ denotes the atom density and $n_D$ denotes the density of dimers.
The rate constant $\beta$ can also be obtained using the optical theorem:
\begin{equation}
  \beta=-\frac{6\pi}{m}{\rm Im}\; a_{\rm AD}~,
\end{equation}
where $a_{\rm AD}$ denotes the atom-dimer scattering length:
\begin{equation}
  \label{eq:betaT0}
  \beta=\frac{20.3\sinh(2\eta_*)}{\sin^2(s_0\ln(a/a_*))
+\sinh^2\eta_*}\frac{ a}{m}~,
\end{equation}
where $a_*$ denotes the value of the two-body scattering length at which there
is a trimer at the atom-dimer threshold: $a_*\approx
0.0798~\kappa_*^{-1}$. This expression was first derived in
\cite{Braaten:2003yc} using Efimov's radial law discussed in the previous
section.

Atom-dimer relaxation at finite temperature was first considered by Braaten
and Hammer in Ref.~\cite{Braaten:2006nn}. Recently, Helfrich and Hammer
\cite{Helfrich:2009uy} extended this work. Using generalized Bose-Einstein
distribution functions to perform a thermal average, they calculated the rate
of change in the number of shallow dimers $N_D$ due to relaxation into deep
dimers:
\begin{equation}
\label{eq:beta_T}
  \frac{\hbox{d}}{\hbox{d}t}N_D=-\int\hbox{d}^3 r \hbox{d}^3p_A \hbox{d}^3p_D
n_A(p_A,r) n_D(p_D,r) \frac{3k}{2m}\sigma_{AD}^{\rm inel.}~,
\end{equation}
here $k=|2\pvec_A-\pvec_D|/3$ denotes the relative momentum,
$\pvec_A$ and $\pvec_D$ are the momenta of atoms and dimer, respectively. The 
generalized Bose-Einstein distribution functions $n_{A/D}$ for atoms/dimers in
a harmonic trapping potential are
\begin{equation}
  \label{eq:bose-einstein-distr}
  n_i(p_{i},r)=\left\{\exp\left[\left(\frac{p_i^2}{2m_{i}}
+\frac{m_{i}\omega^2 r^2}{2}-\mu_{i}\right)/(k_B T)\right]-1 \right\}^{-1}~,
\end{equation}
\begin{figure}[t]
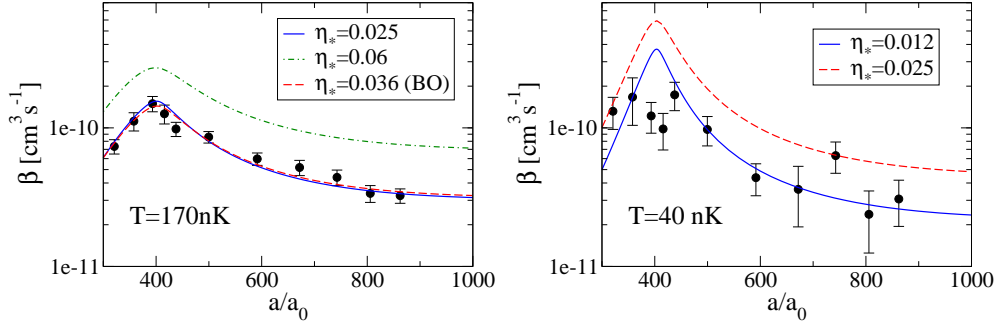

	\centerline{
\includegraphics*[width=6.5cm]{bnew170.eps}
\includegraphics*[width=6.5cm]{bnew40.eps}}
\caption{Left: The dimer relaxation coefficient $\beta$ as a function of 
$a/a_0$ for $T=170$~nK, $a_*=395 a_0$, and different values of 
$\eta_*$. 
BO indicates a Boltzmann average.
Right: $\beta$ as a function of
$a/a_0$ for $T=40$~nK, $a_*=395 a_0$, and different values of
$\eta_*$. The data points in both panels are from~\cite{Knoop08}.
\label{fig:relaxation}}
\end{figure}
with $i=A,D$, $\omega$ denotes the average trap frequency, and $\mu_i$ is the
chemical potential. Knowledge of
the $S$-wave phase shift for atom-dimer scattering $\delta_0^{AD}$ allows
therefore the evaluation of Eq.~(\ref{eq:beta_T}) since $\sigma_{AD}^{\rm
  inel.}=\sigma_{AD}^{\rm total}-\sigma_{AD}^{\rm elastic}$ and
\begin{subequations}
\begin{eqnarray}
\nonumber
  \frac{\hbox{d}\sigma_{AD}^{\rm elastic}}{\hbox{d}\Omega}&=&\left| \frac{1}{k
  \cot\delta_0^{AD}-i k}\right|^2~,\\
\sigma_{AD}^{\rm total}&=&\frac{4\pi}{k}{\rm Im}\left(\frac{1}{k\cot\delta_0^{AD}-ik}\right)~,
\end{eqnarray}
\end{subequations}
Braaten and Hammer have calculated the phase shift $k a\cot\delta_0^{AD}$ for
$k$ up to the dimer breakup threshold $\sqrt{3}/(2a)$
\cite{Braaten:2003yc}. They parameterized the results as
\begin{eqnarray}
\label{phaseshift}
ka \cot\delta_0^{AD}(k)&=&c_1(ka)
+c_2(ka) \cot\bigl[s_0\ln\bigl(0.19\, a/a_*\bigr)+\phi(ka)\bigr]\,,
\end{eqnarray}
where
\begin{eqnarray}
c_1(ka)& = &-0.22+0.39k^2a^2-0.17k^4a^4\,, \nonumber\\
c_2(ka)& = &0.32+0.82k^2a^2-0.14k^4a^4\,, \nonumber\\
\phi(ka)& = &2.64-0.83k^2a^2+0.23k^4a^4\,.
\label{eq:kcotdpara}
\end{eqnarray}
Using these results Helfrich and Hammer evaluated the integral in
Eq.~(\ref{eq:beta_T}). They related there results to the experimental results
by Knoop {\it et al.}~\cite{Knoop08} using the loss model employed in
\cite{Knoop08}. They find
\begin{equation}
\beta \equiv -\frac{\sqrt{27}}{8 \bar{n}_A N_D} \frac{d}{dt}N_D\,.
\end{equation}
The factor of $(\sqrt{3}/2)^3$ takes into account the assumed Boltzmann
distributions $\exp(-m_i\omega^2 r^2/3)$ of the atoms and the dimers that were
assumed in the analysis of Ref.~\cite{Knoop08}.

The free parameters $a_*$ and $\eta_*$ are fitted to the data at $T=170$~nK.
In the left panel of Fig.~\ref{fig:relaxation}, the results for $a_*=395 a_0$
and $\eta_*=0.025,0.06$ (solid curves) give an excellent fit to the
experimental results by Knoop {\it et al.}~\cite{Knoop08}. The dashed curve in
the left panel is the result one obtains by using a Boltzmann instead of a
Bose-Einstein distribution in Eq.~(\ref{eq:bose-einstein-distr}). The results
display some sensitivity to $\eta_*$. At $T=40$~nK, the best fit is obtained
with $\eta_*=0.012$ (solid line). The value $\eta_*=0.025$, which gave the
best fit for 170~nK gives a prediction that is too large by a factor of two
when compared to the experimental results.
\subsection{Four-Body Recombination}
\label{subsec:4reco}
Stecher, d'Incao and Greene considered in their work on the four-body system
also the implications of 4-body universality on loss rates in systems of
ultracold atoms~\cite{Stecher:2009}. In particular, they pointed out that the
existence of the two universal tetramer states discovered by Platter and Hammer
should lead to observables loss features at values of the scattering length
that are related to the scattering length at which the tetramer binding energy
becomes zero. Their results for the values of the scattering lengths are given
in Eq.~(\ref{eq:tetra-scatteringlengths}).

In a second paper Stecher, d'Incao and Greene addressed the problem of
dimer-dimer collisions~\cite{Stecher:2009} which is relevant for the case of
positive scattering length.  They denoted the two-body scattering length for
which the $i$th four-body bound state crosses the dimer-dimer threshold by
$a^*_{dd,i}$. They denoted the two-body scattering length at which the
trimer-atom and dimer-dimer channels become degenerate by $a_{dd}^c$. We
previously defined $a_*$ to be the scattering length at which the trimer state
crosses atom-dimer threshold. The ratio of these quantities are universal
numbers
\begin{equation}
  a^*_{dd,0}/a^c_{dd}\approx 2.37~,\quad a^*_{dd,1}/a^c_{dd}\sim~6.6~,
\quad{\rm and}\quad a^c_{dd}/a_*\approx 6.73~.
\end{equation}
Ferlaino {\it et al.} recently studied the four-body problem with
short-range interactions experimentally~\cite{Ferlaino:2009}. Using ultracold
$^{133}$Cs atoms in the lowest hyperfine state at a temperature of
50~nK, they found  loss features at scattering lengths $-730 a_0$ and $-410 a_0$ which were
interpreted as the four-body loss features predicted by Stecher, d'Incao and
Greene \cite{Stecher:2008}. With the triatomic Efimov resonance measured at
$-870 a_0$, this gives for the ratios of the four- and three-body resonance
position
\begin{equation}
    a^*_{4,0}/ a_* \approx 0.47\quad{\rm and}\quad a^*_{4,1}/ a_*\approx 0.84~.
\end{equation}
These experimental results are in fact surprisingly close to the zero-range
prediction made in \cite{Stecher:2008} since finite range effects are expected
to important at these values of the scattering length. The range of the Cs-Cs
interaction (which is set by the van-der Waals length scale) is approximately
$200 a_0$.

\subsection{Challenges and Opportunities I}
\label{sec:amo-challenges}
Recent experimental progress demonstrates that the limits of complexity have
not been reached yet in the field of few-body dynamics in gases of ultracold
atoms. The experimental evidence for three-body universality in systems
of ultracold atoms has increased significantly. First experimental evidence for
four-body universality have been presented and more can expected in the near
future.

Measurements of Efimov loss features in heterogenous systems, i.e. systems
with different constituent masses, have also been presented. It is therefore
a natural task to identify the universal signatures and their relation to each
other using the short-range EFT. Some of these signatures might offer also
stronger signatures of Efimov physics since the discrete scaling factor
associated with the Efimov bound state spectrum depends on the mass ratios of
the constituents.

Significant progress in mapping out the relevant signatures of universal
four-body physics has been made. However, a full treatment of scattering
processes remains desirable. This would facilitate an analysis of the impact
of universality on further quantities such as the dimer-dimer or atom-trimer
scattering lengths. Access to scattering quantities is also required for the
calculation of the four-body recombination rate at finite temperature.

The ongoing analysis of the effects of range corrections on universal few-body
physics is important since most experiments are carried out at scattering
lengths at which the finite range of the two-body interaction is expected to
have a measurable effect. The questions that remain to be answered are
therefore how universal relations that relate different recombination
observables will be affected by finite range effects and what implications
does a small but finite range have for the four-body bound state spectrum.
Since the linear range correction has been shown to be small for some
observables it will be important to address the quadratic range correction
with a full N2LO analysis.

\section{Low-Energy Universality in Nuclear Physics}
\label{sec:nuc}
The model-independent description of nuclear systems has been one of the main
goals in nuclear physics for many years. Since their introduction to nuclear
physics, EFTs have allowed the calculation of a large number of observables
model-independently and in some cases to very high accuracy.  The {\it standard} EFT
approaches to internucleon interactions employ nucleon and pion fields as the
minimal set of degrees of freedom
\cite{Epelbaum:2005pn,Bedaque:2002mn,Epelbaum:2008ga}. The separation of
scales that is exploited here is the one of chiral symmetry which identifies
pions as the Goldstone bosons of the spontaneously broken chiral symmetry of
QCD. The relatively small pion mass is then a consequence of the small up and
down quark mass, i.e. the explicitly broken chiral symmetry.

Fortunately, this is not the only separation of scales in the two-nucleon
system. It turns out that also the scattering length in the nucleon-nucleon
system is large compared to the range of the interaction which in this case is
set by the pion mass $m_\pi$. The short-range EFT can therefore be applied in
the nuclear sector as long as we consider momenta that are much smaller than
$m_\pi$.

Such a limitation might seem to make this approach inadequate for nuclear
physics. However, it is important to realize that a large number of
reactions relevant to nuclear astrophysics occur at energies well below this
breakdown scale of the {\it short-range} EFT.  Proton-proton fusion
($p+p\rightarrow\: ^2$H$+e^+ + \nu_e$) for example, a reaction which plays an
important role in the sun's energy generation, occurs at the keV scale due to
the {\it low} temperature in the sun. The short-range EFT seems therefore to
be the ideal framework for the calculation of such reaction.

Halo nuclei and weakly bound systems of $\alpha$ particles are an additional
playground for the short-range EFT. Halo nuclei are weakly bound systems of a
core (e.g. an $\alpha$-particle) and additional nucleons that are weakly bound
to the core. There is also evidence that some of the properties of
$\alpha$-clusters can be described using the short-range EFT.

In this section we will discuss applications of the short-range EFT to systems
of nucleons. We will first consider the two-nucleon system and discuss a
recent calculation in this sector which exemplifies the strong predictive
power of the short-range EFT. We will then turn to the three-nucleon sector.
We will discuss the renormalization, the inclusion of higher order corrections
and the status of the inclusion of external currents in the few-body
sector. Then we summarize recent calculation of halo nuclei. We end with a
subsection which discusses the work required to be done in the future.

\subsection{Two Nucleons}
Nucleons are spin 1/2 particles that come in two flavors, protons and
neutrons.  A nucleon field will therefore carry spin and isospin indices to
accommodate these extra degrees of freedom. The Lagrange density for nucleons
interacting at very low energies is constructed by writing down all possible
operators allowed by the underlying symmetries
\begin{eqnarray}
  \label{eq:Lag3N}
  {\cal L}&=&N^\dagger \left(i\partial_0 +\frac{\vec{\nabla}^2}{2m}\right)N
\nonumber 
-C_0^t\left(N^T \tau_2 \sigma_i \sigma_2 N\right)^\dagger\left(
N^T \tau_2 \sigma_i \sigma_2 N\right) 
\\
& &\hspace{3cm} -C_0^s\left(N^T \sigma_2 \tau_a \tau_2 N\right)^\dagger\left(
N^T \sigma_2 \tau_a \tau_2 N\right) +...\,,
\end{eqnarray}
where the dots represent higher-order contributions suppressed by more fields
and/or derivatives. The low-energy constants $C_0^s$ and $C_0^t$ are
renormalized to the spin-singlet and triplet scattering length, respectively.

The short-range EFT has been applied very successfully in the two-nucleon
sector to electroweak observables. Here, we will only mention a few very
recent calculations since they exemplify typical applications and the
predictive power for low-energy processes with only a few physical input
parameters.

Christlmeier and Grie\ss hammer~\cite{Christlmeier:2008ye} recently addressed
a discrepancy between previously obtained theoretical and experimental results 
for the electro-disintegration cross section of deuterium ($^2$H$(e,e'p)n$).
They considered the triple-differential cross-section for this process which
can also be decomposed into
\begin{eqnarray}
\nonumber
  \frac{\dd^3\sigma}{\dd E_e^{\mathrm{lab}}\,\dd\Omega_e^{\mathrm{lab}}\,
    \dd\Omega_p} &=&
\frac{\dd^3}{\dd E_e^{\mathrm{lab}}\,\dd\Omega_e^{\mathrm{lab}}\,\dd\Omega_p} 
  (\sigma_L+\sigma_T\\
&&\hspace{3cm}+\sigma_{LT} \cos \Phi_p + \sigma_{TT} \cos 2\Phi_p)\;\;,
  \label{decomp}
\end{eqnarray}
where $\Omega_e^{\mathrm{lab}}=(\Theta_e^{\mathrm{lab}},
\Phi_e^{\mathrm{lab}}\equiv0)$ and $\Omega_p=(\Theta_p,\Phi_p)$ are the
scattering angles of electron and proton in the lab frame, respectively. The
superscript ``lab'' denotes  the laboratory frame which is defined to be the
rest frame of the deuteron.
They found their results to be in excellent agreement with previous
theoretical calculations by Arenh\"ovel {\it et al.} \cite{ArenLeideTom04} and
Tamae~\cite{Tamae} which lead to the conclusion that none of the available
theoretical approaches can explain the experimental data presented for
$\sigma_{LT}$ in Ref.~\cite{sdalinac}.
A subsequent experimental study of the double differential cross section for
($^2$H$(e,e'p)n$) at an angle of $180^\circ$ lead to excellent agreement between
experiment and theory \cite{Ryezayeva:2008zz}.

Finally, Ando considered $p p \rightarrow  pp\pi^0$ near production threshold
in the short-range EFT~\cite{Ando:2007in} and Phillips, Schindler and Springer
considered parity violating nucleon-nucleon scattering in \cite{Phillips:2008hn}.
\subsection{Three Nucleons}
The short-range EFT was first applied to three-nucleon systems by Bedaque,
Hammer and van Kolck \cite{Bedaque:1998mb,Bedaque:1999ve}.  As in the bosonic
case, the calculation of three-body processes is simplified if we introduce
auxiliary fields that carry the quantum numbers of the allowed two-body
$S$-wave states.  Here, we introduce two auxiliary fields (one for every
possible $S$-wave channel), $t_i$ with spin 1 (isospin 0) and $s_i$ with spin 0
(isospin 1), respectively
\begin{eqnarray}
\nonumber
\mathcal{L}&=&N^\dagger(i\partial_0+\frac{\nabla^2}{2M})N
-t_i^\dagger(i\partial_0+\frac{\nabla^2}{4M}-\Delta_t)t_i
-s_j^\dagger(i\partial_0+\frac{\nabla^2}{4M}-\Delta_s)s_j\\ \nonumber
&& \quad+g_t \left( t_i^\dagger N^T \tau_2 \sigma_i \sigma_2
N+\text{h.c}\right) +g_s \left( s_j^\dagger N^T \sigma_2 \sigma_j
\tau_2 N+\text{h.c}\right) \\
\nonumber
&&\hspace{10mm}-G_3
N^\dagger\biggl(g_t^2(t_i\sigma_i)^\dagger t_{i'}\sigma_{i'}
+\frac{1}{3}g_t g_s[(t_i\sigma_i)^\dagger s_j\tau_j+\text{h.c}]
\\
&&\hspace{60mm}+g_s^2(s_j\tau_j)^\dagger s_{j'}\tau_{j'}\biggl)N+\ldots~,
\label{eq:lagrange3N}
\end{eqnarray}

We want to use this Lagrange density to calculate observables for nucleon
deuteron scattering. The total spin of this system is either 1/2 (singlet) or
3/2 (quartet) and correspondingly two different integral equations can be
derived that describe the scattering of a neutron and a deuteron in a relative
$S$-wave. The integral equation for scattering in the quartet channel is given
by
\begin{equation}
  \label{eq:quartet}
  t_{3/2}(p,k)=-\frac{4\pi\gamma_t}{m}K(p,k)-\frac{1}{\pi}\int_0^\infty\hbox{d}q\,
  q^2 D_s(q) K(p,q)t_{3/2}(q,k)~,
\end{equation}
where
\begin{equation}
  K(p,k)=\frac{1}{p k}\ln\left(\frac{p^2+p k+k^2}{p^2-pk+k^2}\right)~.
\end{equation}
The Pauli principle forbids the appearance of an $S$-wave three-body force in
this channel, since the spins and isospins of two of the three nucleons are
aligned parallely.

The integral for the scattering in the singlet channel is given by
\begin{eqnarray}
  \label{eq:doublet}
\nonumber
  t^s_{1/2}(p,k)&=&\frac{8\pi \gamma_t}{m}\left(\frac{3}{4}K(p,k)+\frac{2H}{\Lambda^2}\right)
+\frac{1}{2\pi}\int_0^\Lambda\hbox{d}q^2 D_s(q)(K(q,p)+\frac{2H}{\Lambda^2}) t_s(q,k)\\
\nonumber
&&\hspace{3cm}+\frac{1}{2\pi}\int_0^\Lambda\hbox{d}q^2 D_t(q)(K(q,p)+\frac{2H}{\Lambda^2}) t_s(q,k)\\
\nonumber
t^t_{1/2}(p,k)&=&\frac{8\pi \gamma_t}{m}\left(\frac{1}{4}K(p,k)+\frac{2H}{\Lambda^2}\right)+
\frac{1}{2\pi}\int_0^\Lambda\hbox{d}q^2 D_t(q)(K(p,q)+\frac{2H}{\Lambda^2}) t_s(q,k)\\
\nonumber
&&\hspace{3cm}+\frac{1}{2\pi}\int_0^\Lambda\hbox{d}q^2 D_s(q)(K(p,q)+\frac{2H}{\Lambda^2}) t_s(q,k)~.\\
\end{eqnarray}
In this case a three-body force is allowed and also required for consistent
renormalization of the problem. The triton binding energy can be calculated
from the homogeneous part of Eq.~(\ref{eq:doublet}) since the total spin
(isospin) of the triton is 1/2 (1/2).

The solutions of Eqs.~(\ref{eq:quartet}) and (\ref{eq:doublet}) are related to
the deuteron-neutron phase shift
\begin{subequations}
\begin{eqnarray}
  t_{3/2}(k,k)&=&\frac{3\pi}{m}\frac{1}{k\cot\delta_{3/2}-ik}~,\\
  t^t_{1/2}(k,k)&=&\frac{3\pi}{m}\frac{1}{k\cot\delta_{1/2}-ik}~.
\end{eqnarray}
\end{subequations}
The appearance of a three-body force in the doublet channel implies the
correlation of different three-body observables.  The short-range EFT
describes these correlations therefore with the minimal number of degrees of
freedom and all observed correlation lines are therefore a consequence of the
large scattering length in the two-nucleon system.  The short-range EFT also
offers a different perspective on the internucleon interaction: different
two-nucleon interactions may describe two-body data equally well but give
different predictions for three-body observables.

\paragraph{\it Finite Range Corrections:\\}
Since the ratio of effective range over scattering length in the spin-triplet
channel is approximately
$R/a\sim 1/3$, the consistent calculation of finite range corrections is crucial
for an accurate description of low-energy observables in nuclear physics.
\begin{figure}[t]
\centerline{\includegraphics*[width=10cm,angle=0]{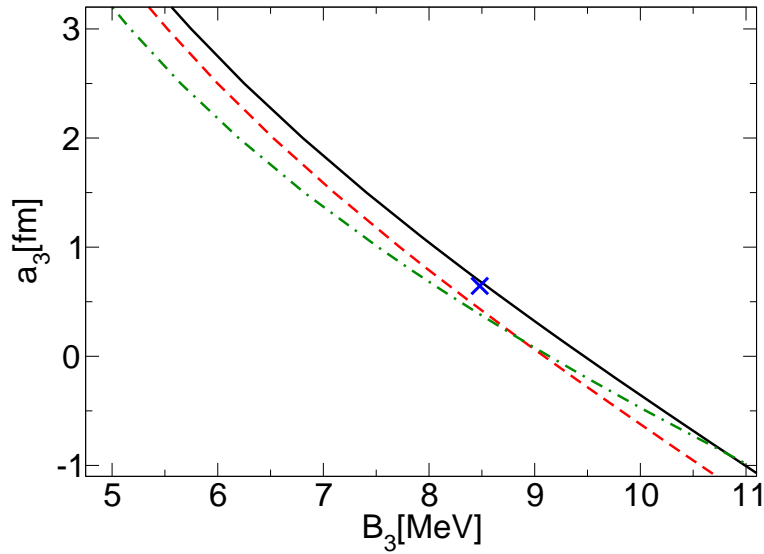}}
\caption{\label{fig:phillipsline}The Phillips line for leading (dot-dashed
  line), next-to-leading (dashed line) and next-to-next-to-leading
(solid line) order. The cross denotes the experimental value.}
\end{figure}
In the three-nucleon context, range corrections were first addressed by Hammer
and Mehen \cite{Hammer:2001gh}. They included the the first subleading
correction perturbatively.  Bedaque {\it et al.} \cite{Bedaque:2002yg}
calculated finite range corrections up N2LO in the EFT expansion using the
modified version of the STM equation discussed in subsection
\ref{subsec:3body}. They calculated phase shifts for neutron-deuteron
scattering up to N2LO using the triton binding energy and the neutron-deuteron
scattering length as three-body input parameters. In \cite{Platter:2006ad} the
powercounting results from Ref.~\cite{Platter:2006ev} were used to calculate
scattering phase shifts for $S$-wave scattering in the neutron-deuteron
doublet channel up to N2LO. The triton binding energy was calculated as a
function of the neutron-deuteron scattering length $a_3$.  In
Fig.~\ref{fig:phillipsline} we show the result for this correlation line (the
Phillips line).  The dot-dashed, dashed and solid line show the LO, NLO and
N2LO results, respectively. The cross denotes the experiment result for the triton
binding energy and the neutron-deuteron scattering length. The lines in
Fig.~\ref{fig:phillipsline} are not parallel to each other since the shift
from order to order depends on two expansion parameters, $\kappa r_s$ and
$\gamma r_s$ (with $\kappa=\sqrt{m B_t}$).  At the experimental value of the
$nd$ scattering length we find for the triton binding energy
\begin{equation}
  \label{eq:B_triton}
  B_t=(8.08({\rm LO})+0.11({\rm NLO})+0.35({\rm N2LO}))~\hbox{MeV}~.
\end{equation}
These $8.54$~MeV are in very good agreement with the experimental value
$B_t^{exp}=8.48$~MeV and within the naively expected error bounds of a N2LO
calculations.  These results furthermore exemplify the unusual convergence
pattern of the short-range EFT for three-body observables. This is due to the
unnaturally small NLO correction while the N2LO corrections has the size as
predicted by naive dimensional analysis. We explained in Section
\ref{subsec:3body} that in the unitary limit the discrete scale invariance of
the LO wave function prohibits the bound state spectrum to obtain any NLO
correction in the unitary limit. It is tempting to speculate that the
approximate discrete scale invariance of the leading order amplitude at
unitarity is also partially responsible for the unexpected convergence pattern
in the three-nucleon sector.

\paragraph{\it Electroweak Observables:\\}
A major benefit of using a field-theoretic framework is the straightforward
inclusion of external currents. The short-range EFT facilitates therefore the
calculation of electroweak observables of few-nucleon reactions.  This is
particularly useful since a number of reactions relevant to nuclear
astrophysics occur at energies well below the breakdown scale of the
short-range EFT which is set by the pion mass.  The number of three-body
calculations with external currents is however, extremely limited.  Universal
properties of the electric form factor and charge radius of the triton were
considered in \cite{Platter:2005sj}. The leading order wave function was used
to evaluate the charge form factor from the leading order charge operator
\begin{equation}
\label{eq:ffdef}
F_C({\bf q}^2)=\langle\Psi_{{\bf K}+{\bf q}} \;{\bf k}_f|\rho_C|{\bf k}_i\;
\Psi_{\bf K}\rangle~,
\end{equation}
where ${\bf q}={\bf k}_i -{\bf k}_f$, ${\bf k}_i$ and ${\bf k}_f$ 
are the initial and final momentum of the scattered electron, and
$\Psi_{\bf K}$ denotes the full triton wave function with center of mass 
momentum ${\bf K}$. 
The charge density operator $\rho_C$ is defined as \cite{Friar81}
\begin{equation}
\rho_C=\sum^3_i\Bigl[\frac{1}{2}(1+\tau_{iz})\rho_C^p({\bf r}-{\bf r}_i)
+\frac{1}{2}(1-\tau_{iz})\rho_C^n({\bf r}-{\bf r}_i)\Bigr]~.
\end{equation}
The charge radius $\langle r^2 \rangle$ can then be defined as 
\begin{equation}
  \label{eq:chrad}
F_C({\bf q^2})= 1 - {\bf q^2} \langle r^2 \rangle/6 +\ldots\,.
\end{equation}
\begin{figure}[tb]
\centerline{\includegraphics*[width=10cm,angle=0]{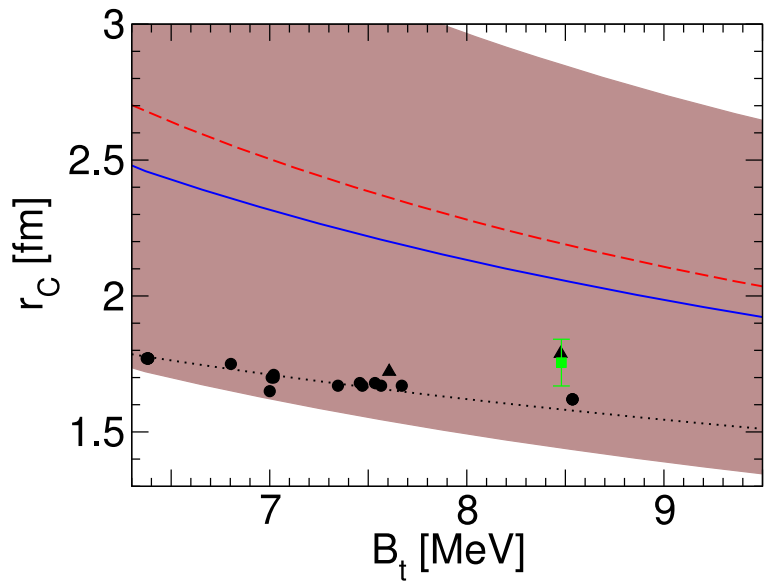}}
\caption{\label{fig:radius3H}The correlation between the triton
charge radius and binding energy. 
The solid (dashed) line denotes the leading-order
result using $a_t$ and $a_s$ ($B_d$ and $a_s$) as input parameters.
The circles indicate Faddeev calculations using different 
potentials from Ref.~\cite{Chen85} while the square gives the experimental
values.
}
\end{figure}
In Fig.~\ref{fig:radius3H} we display the one-parameter correlation
between the charge radius and the triton binding energy. The two lines
shown in this figure correspond to two different inputs for the two-body
triplet scattering length. The shaded band gives the expected error of a
leading order calculation. The existence of this correlation was (to our
knowledge) first pointed out in Ref.~\cite{Friar85}.

A process relevant to big bang nucleosynthesis is thermal proton capture
($p+d\rightarrow^3$He$+\gamma$). The calculation of this amplitude is
complicated by the presence of Coulomb effects. Yet, a first step towards the
goal of calculating such processes within the short-range EFT framework was
performed in \cite{Sadeghi:2006aa,Sadeghi:2005aa}. In this work the authors
considered thermal neutron capture ($n+d\rightarrow\:^3$H$+\gamma$) and calculated the
total cross section at zero energy up to N2LO. The calculation of cross
sections for this process is easier since no Coloumb effects need to be
considered in this reaction. The final result is
\begin{eqnarray}
  \label{eq:iran}
\nonumber
  \sigma_{\rm
    tot}&=&\left[0.485(\hbox{LO})+0.011(\hbox{NLO})+0.007(\hbox{N2LO})\right]~\hbox{mb}
\\
&=&\left[0.503\pm 0.003\right]~\hbox{mb}~,
\end{eqnarray}
where the remaining uncertainty is an estimate of higher order effects. This
theoretical result is in excellent agreement with the experiment value
$\sigma_{\rm tot}^{\rm Exp}=0.508\pm 0.015$~mB.

 A first calculation of Coulomb effects in
proton-deuteron scattering (in the quartet channel) was done in
\cite{Rupak:2001ci}. However, the energies relevant to big bang
nucleosynthesis are significantly lower than considered in this publication
and still a challenge.

\subsection{Four and more Nucleons}
The short-range EFT has also been applied to the four-nucleon sector
\cite{Platter:2004zs}. The binding energy of the $\alpha$-particle
was calculated using the same approach as discussed in
subsection \ref{subsec:4atoms}. In the case of the nuclear system,
two effective potentials have to be introduced:
\begin{eqnarray}
  V_s&=&\lambda_s\mathcal{P}_s|g\rangle\langle g|~,\\
V_t&=&\lambda_t\mathcal{P}_t|g\rangle\langle g|~,
\end{eqnarray}
where $V_s$ ($V_t$) denotes the spin-singlet (triplet) potential.
$\mathcal{P}_s$ and $\mathcal{P}_t$ projects on the spin-singlet and triplet
channels, respectively.  The coupling constants $\lambda_s$ and $\lambda_t$ are
therefore renormalized to the singlet and triplet scattering lengths.
The effective three-body potential is 
\begin{equation}
  V_3=\lambda_3\,\mathcal{P}_A\,|\xi\rangle\langle \xi|~,
\end{equation}
where $\mathcal{P}_A$ is the operator that projects on the completely
antisymmetrized three-nucleon state. The three-body coupling constant
$\lambda_3$ can be adjusted using the triton binding energy.

\begin{figure}[tb]
\centerline{\includegraphics*[width=10cm,angle=0]{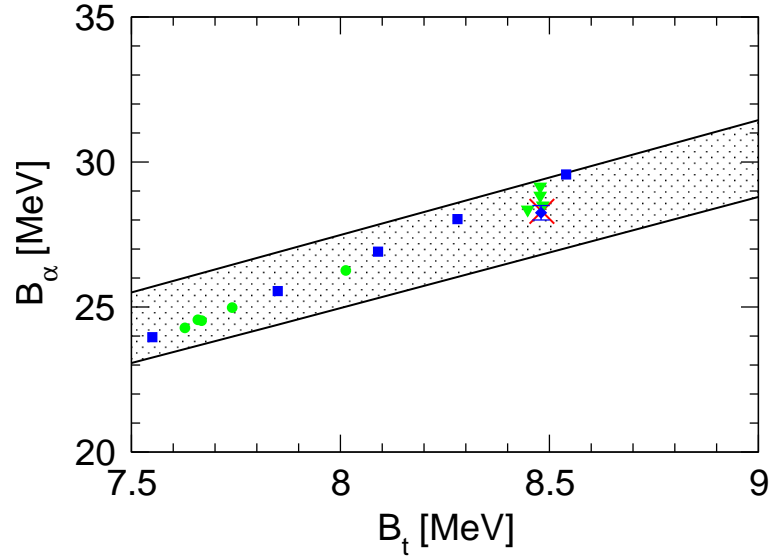}}
\caption{\label{fig:tjon} The correlation between triton and alpha particle
  binding energies, as predicted by the short-range EFT. The grey circles and
  triangles show various calculations using phenomenological potentials
  \cite{Nogga:2000uu}.  The squares show the results of chiral EFT at NLO for
  different cutoffs while the diamond gives the N$^2$LO result
  \cite{Epelbaum:2000mx,Epelbaum:2002vt}.  The cross shows the experimental
  point.  }
\end{figure}

The correlation line between triton and the $\alpha$ particle binding energy
were calculated using the Faddeev and Faddeev-Yakubovski equations.  In
Fig.~\ref{fig:tjon} we show this correlation which is also known as the Tjon
line.
Since the short-range EFT uses the minimal number of degrees of freedom for
the construction of the interaction, it explains therefore the old observation
of the linear correlation between $B_t$ and $B_\alpha$. In Fig.~\ref{fig:tjon}
we show for example also calculations using phenomenological potentials
\cite{Nogga:2000uu} and a chiral EFT potential with explicit pions
\cite{Epelbaum:2000mx,Epelbaum:2002vt}.  All these calculations have to lie in
the band generated with the short-range EFT since these potentials reproduce
the large scattering length of the internucleon interaction. By the same
logic, the short-range EFT also explains why different versions of the
renormalization group evolved potential $V_{\rm low\: k}$ reproduce the Tjon
line \cite{Nogga:2004ab}.

The short-range EFT was also used to calculate the properties of heavier
systems than the $\alpha$-particle. Stetcu {\it et al.} \cite{Stetcu:2006ey}
considered the $\alpha$-particle and $^6$Li using the short-range EFT
in combination with the no-core shell model. They reproduced the ground-state
results for the $\alpha$-particle and gave an estimate for the binding energy
of its first excited 0$^+$ state. The agreement with the experimental value
of the $^6$Li ground state was at the 70 \% level.

\subsection{Halo Nuclei}
Another exciting arena for the short-range EFT is the physics of halo nuclei.
Halo nuclei are weakly bound systems which consist of a tightly bound core and
a ``halo'' of one or more nucleons. The radius of a halo nucleus is typically
much larger than the radius of the core since the total binding energy of such
a system is comparable to the binding energy of its core. A well-known example
is $^6$He, a system consisting of an $\alpha$-particle core and two neutrons.

Over the last years ab-initio, wave functions methods (that start from a
realistic internucleon interaction) have made considerable progress in
describing the lighter of the known halo systems \cite{Bacca:2009yk}, however
calculations for heavier systems such as $^{11}$Li or $^{20}$C are not under control
yet. The short-range EFT offers an alternative perspective on these systems by
treating the core of the halo as a separate degree of freedom. A system such
as $^6$He becomes therefore an effective three-body problem and it's
properties can be calculated with the methods discussed in the previous
sections. Such an approach is interesting for several reasons. It allows to
test whether these systems have features related to large scattering length
physics such as excited Efimov states but provides also testable predictions
for observables such matter and charge radii.  A further advantage of this
approach is that scattering observables become directly accessible and can be
calculated without any further approximations.

The first application of the short-range EFT to halo nuclei was carried out in
Refs.~\cite{Bertulani:2002sz,Bedaque:2003wa}.  In these works the authors
considered the one-neutron halo $^5$He and calculated in particular phase
shifts and cross sections for elastic $\alpha$-nucleon scattering. A further
example of a nuclear two-body cluster that has been considered is the
2-$\alpha$ system~\cite{Higa:2008dn}.

\begin{figure}[t]
\centerline{\includegraphics*[width=4.in,angle=0]{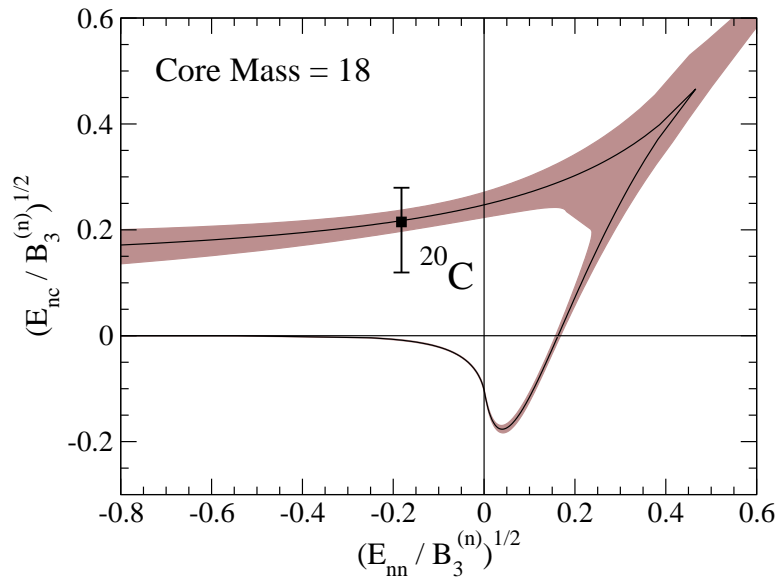}}
\caption{\label{fig:halo} Boundary curve in the $\sqrt{E_{nc}/B_3^{(n)}}$
  vs. $\sqrt{E_{nn}/B_3^{(n)}}$ plane with leading order error bands.
  Boundary curve shown for a core mass of $A = 18$ with the experimental data
  for $^{20}$C from Ref.~\cite{TUNL}.}
\end{figure}

Recently, Canham and Hammer~\cite{Canham:2008jd} performed the first EFT
calculation for two-neutron halos. In their work they calculated the binding
energies and radii of halos such $^{11}$Li and $^{20}$C. Canham and Hammer
also addressed the question whether any of the considered systems supports an
excited Efimov state.  Fig.~\ref{fig:halo} shows a parametric plot
($(E_{nc}/B_3^{(n)})^{1/2}$ versus $(E_{nn}/B_3^{(n)})^{1/2}$) which describes
the region in the two-body parameter space that supports a three-body state
above $B_3^{(n)}$. They found that the $^{20}$C system might exhibit an
excited Efimov state close to the threshold.
The $^6$He halo nucleus was not considered by Canham and Hammer since it is
expected to have $P$-wave resonance in the neutron-core subsystem.  This
feature raises also the question whether a three-body force is required for
consistent renormalization in this system. A successful description of the
effective three-body system would furthermore lead to the exciting problem of
describing $^8$He (the effective five-body problem) within the short-range
EFT.
\subsection{Challenges and Opportunities II}

Currently, calculations for observables have only been performed up to N2LO in
the EFT expansion. A computation up to N3LO would provide further information
on the convergence radius of the short-range EFT. This is also required to
fulfill the original promise of outstanding accuracy for few-body observables
in the few-nucleon sector.

The inclusion of Coulomb effects has also to be addressed in the near
future. It is known in the two-body case how to calculate the scattering
of, for example, two protons. This problem has not been solved to full
satisfaction in the three-body case although it is crucial for the calculation
of scattering processes relevant to nuclear astrophysics such $pd\rightarrow\:
^3$He$\:\gamma$. Progress in this direction is also required for further
applications of the short-range EFT in the four-nucleon sector such as a
calculation of the cross sections for the processes commonly denoted as HEP
($^3$He$+p\;\rightarrow\;^4$He$+e^++\nu_e$) and HEN
($^3$He$+n\;\rightarrow\;^4$He$+\gamma$).

The calculation of such four-body processes will furthermore require the
development of numerical tools which facilitate the calculation of the
corresponding scattering amplitudes\footnote{While this manuscript was
  finalized we became aware of a first calculation of four-nucleon scattering
  observables using the framework of the short-range
  EFT~\cite{Kirscher:2009aj}.}.  With such tools at hand, the existence of
further universal correlation lines could be explored (e.g. the correlation
between nucleon-triton scattering length and binding energy as presented for
example in \cite{Deltuva:2006sz}).

\section{Final Words}
\label{sec:summary}
We have demonstrated that the short-range EFT is an excellent tool to analyze
the properties of systems whose constituents exhibit a large two-body
scattering length.

Atomic physics provides a good testing ground and gives additional
justification to analyze the more general (scattering length dependent)
implications for such systems. It furthermore gives challenges to find
solutions to problems that are not encountered in the nuclear physics case
such as, for example, deeply bound two-body states that need to be accounted for
in a sound manner.

The short-range EFT has also provided an alternative perspective on the topic
of three-body forces. The importance of three-body forces depends strongly on
the resolution at short distances of the chosen approach. In EFT approaches to
nuclear systems that employ pionic degrees of freedom, the first three-body forces are generally
required at higher orders in the EFT expansion. There is therefore nothing
fundamental about a three-body force, but it is rather a further necessary tool
to account for unknown short-distance physics.

It has also been shown that pionic degrees of freedom are irrelevant to the
description of a number of important quantities such as the triton bound state
or the $\alpha$-particle. A number of observables have been computed beyond
leading order and excellent agreement with experiment and/or calculations
using realistic internucleon was found.

We have tried to supply a number of open problems that should be addressed in
the near future. We hope they are understood not as barriers but as
exceptional chances for the short-range EFT to largely extend its current
region of applicability.
\begin{acknowledge}
  I thank Eric Braaten, Hans-Werner Hammer and Daniel Phillips for useful
  discussions and valuable comments on this manuscript.  This work was
  supported in part by the National Science Foundation under Grant
  No.~PHY--0653312, and the UNEDF SciDAC Collaboration under DOE Grant
  DE-FC02-07ER41457.
\end{acknowledge}

\end{document}